\documentclass[aps,prd,onecolumn,showpacs,10pt,superscriptaddress,preprintnumbers,nofootinbib,floatfix]{revtex4-1}

\usepackage{amsmath,amssymb,bm}\allowdisplaybreaks[1]
\usepackage{graphicx}
\usepackage{float}
\usepackage{physics, esint}
\usepackage{xcolor}
\usepackage[normalem]{ulem}
\usepackage[colorlinks=true,linkcolor=blue,citecolor=blue,urlcolor=blue]{hyperref}
\usepackage{adjustbox}
\usepackage{tikz}
\usetikzlibrary{arrows.meta}
\usepackage{enumitem}
\usepackage{multirow}
\usepackage{orcidlink}

\DeclareFontFamily{U}{mathx}{\hyphenchar\font45}
\DeclareFontShape{U}{mathx}{m}{n}{<5> <6> <7> <8> <9> <10> <10.95> <12> <14.4> <17.28> <20.74> <24.88> mathx10}{}
\DeclareSymbolFont{mathx}{U}{mathx}{m}{n}
\DeclareFontSubstitution{U}{mathx}{m}{n}
\DeclareMathAccent{\widebar}{0}{mathx}{"73}

\renewcommand{\sec}[1]{Sec.~\ref{#1}}
\newcommand{\fig}[1]{Fig.~\ref{#1}}
\newcommand{\figs}[2]{Figs.~\ref{#1} and~\ref{#2}}
\newcommand{\eq}[1]{Eq.~\eqref{#1}}
\newcommand{\eqs}[2]{Eqs.~\eqref{#1} and~\eqref{#2}}
\newcommand{\refcite}[1]{Ref.~\cite{#1}}
\newcommand{\refs}[1]{Refs.~\cite{#1}}

\newcommand{\pp}[1]{\left(#1\right)}
\newcommand{\bb}[1]{\left[#1\right]}
\newcommand{\cc}[1]{\left\{#1\right\}}
\newcommand{\vv}[1]{\left\langle #1 \right\rangle}
\newcommand{\bigpp}[1]{\big(#1\big)}
\newcommand{\bigbb}[1]{\big[#1\big]}
\newcommand{\bigcc}[1]{\big\{#1\big\}}
\newcommand{\bigvv}[1]{\big\langle #1 \big\rangle}

\newcommand{\Bigvv}[1]{\Big\langle #1 \Big\rangle}
\newcommand{\biggpp}[1]{\bigg(#1\bigg)}
\newcommand{\biggbb}[1]{\bigg[#1\bigg]}
\newcommand{\biggcc}[1]{\bigg\{#1\bigg\}}

\newcommand{\beq}[1][]{\begin{equation}\label{#1}}
\newcommand{\eeq}{\end{equation}}
\newcommand{\bse}[1][]{\begin{subequations}\label{#1}}
\newcommand{\ese}{\end{subequations}}
\newcommand{\nn}{\nonumber}

\def\T{\mathsf T}
\def\S{\mathcal{S}}
\def\ob{{\rm ob}}
\def\sh{{\rm sh}}
\renewcommand{\l}{\ell}

\renewcommand{\O}{\mathcal{O}}

\newcommand{\sgn}{\operatorname{sgn}}

\newcommand{\M}{\mathcal{M}}

\newcommand{\N}{\mathcal{N}}
\renewcommand{\H}{\mathcal{H}}

\newcommand{\GeV}{{\rm GeV}}

\begin{document}

\title{Bayesian reconstruction of pixelated generalized parton distributions}

\author{Zhite Yu\,\orcidlink{0000-0003-1503-5364}}
\thanks{Corresponding author: \href{mailto:zyu1@bnl.gov}{zyu1@bnl.gov}}
\affiliation{High Energy Theory Group, Physics Department, Brookhaven National Laboratory, Upton, New York 11973, USA}
\affiliation{Theory Center, Jefferson Lab, Newport News, Virginia 23606, USA}

\author{Marco Zaccheddu\,\orcidlink{0000-0002-3672-8111}}
\affiliation{Theory Center, Jefferson Lab, Newport News, Virginia 23606, USA}

\author{W. Melnitchouk\,\orcidlink{0000-0002-9521-5973}}
\affiliation{Theory Center, Jefferson Lab, Newport News, Virginia 23606, USA}

\author{Jian-Wei Qiu\,\orcidlink{0000-0002-7306-3307}}
\affiliation{Theory Center, Jefferson Lab, Newport News, Virginia 23606, USA}
\affiliation{Department of Physics, William \& Mary,
Williamsburg, Virginia 23187, USA
\\
\vspace*{0.2cm}
{\bf
JAM Collaboration\\
\vspace*{0.1cm}(GPD Analysis Group)\vspace*{0.2cm}
}}

\author{Nobuo Sato\,\orcidlink{0000-0002-1535-6208}}
\affiliation{Theory Center, Jefferson Lab, Newport News, Virginia 23606, USA}

\date{\today}
\preprint{JLAB-THY-26-4989}

\begin{abstract}
Extracting generalized parton distributions (GPDs) from hard exclusive processes with their full dependence on the parton momentum fraction $x$, skewness $\xi$, and four-momentum transfer squared~$t$, is an important but nontrivial task.
We represent GPDs by pixels on an $(x, \xi, t)$ grid, interpolated with finite elements, to turn their extraction into an exactly solvable linear Bayesian inference problem without parametric bias, having only to specify a prior.
Among the observables tested, the nonscaling integrals of two specific exclusive processes ($\gamma\gamma$ mesoproduction and $\gamma\pi$ photoproduction) provide the strongest constraints, although the GPDs themselves are never uniquely determined within controlled pointwise uncertainties.
However, even when the GPDs are not constrained, one can still interpret existing exclusive scattering data through collinear GPD factorization and make predictions for other hard scattering processes that can be measured in future.
\end{abstract}

\maketitle

\section{Introduction}
\label{sec:intro}

Generalized parton distributions (GPDs) are elementary QCD light-front field correlation functions~\cite{Muller:1994ses, Ji:1996ek, Radyushkin:1997ki} that encode information about the three-dimensional partonic structure of hadrons~\cite{Burkardt:2000za, Burkardt:2002hr}, and provide access to fundamental properties such as energy and angular momentum densities, as well as internal pressure and shear-force distributions~\cite{Ji:1994av, Ji:1995sv, Lorce:2017xzd, Metz:2020vxd, Ji:1996ek, Polyakov:2002yz, Polyakov:2018zvc, Burkert:2018bqq} (see \refs{Ji:1998pc, Goeke:2001tz, Diehl:2003ny, Belitsky:2005qn, Boffi:2007yc} for reviews).
They have attracted considerable theoretical and experimental interest over the past three decades and will remain an important focus of current and future studies of exclusive hard scattering processes at Jefferson Lab and the future Electron-Ion Collider~\cite{Accardi:2012qut, AbdulKhalek:2022hcn, Accardi:2023chb, Accardi:2026slw, Diehl:2023nmm}.

A generic GPD $F(x,\xi,t;\mu)$ at a resolution scale $\mu$ is a function of the light-front momentum fraction~$x$ carried by the active parton, the skewness parameter $\xi$, and the squared momentum transfer $t$ to the hadron.
Although GPDs are well defined in QCD, they cannot yet be fully computed nonperturbatively, requiring efforts to extract them from hard exclusive experimental processes using QCD factorization theorems.
While the variables $\xi$ and~$t$ can be determined directly from experimental kinematics, the variable $x$~enters hard exclusive observables only through convolution integrals using exclusive factorization theorems~\cite{Collins:1996fb, Ji:1998xh, Collins:1998be, Qiu:2022bpq, Qiu:2022pla}.
Consequently, resolving the $x$ dependence of GPDs from experimental data constitutes a nontrivial inverse problem that is fundamentally different from the extraction of parton distribution functions (PDFs)~\cite{Hou:2019efy, Bailey:2020ooq, NNPDF:2017mvq, Alekhin:2017kpj, Accardi:2016qay, H1:2015ubc, Moffat:2021dji, Cocuzza:2026zoy}.

It was recently suggested~\cite{Bertone:2021yyz, Moffat:2023svr} that, with only the well-studied exclusive processes such as deeply virtual Compton scattering (DVCS)~\cite{Ji:1996nm, Radyushkin:1997ki} and timelike Compton scattering (TCS)~\cite{Berger:2001xd}, infinitely many GPD solutions can fit the relevant observables within achievable experimental precision, even with next-to-leading-order and evolution effects included.
A multi-channel analysis, together with input from lattice QCD calculations, has thus been called for.
On the other hand, a number of global GPD analysis programs based on various parametric methods have been carried out~\cite{Kumericki:2007sa, Kumericki:2009uq, Cuic:2023mki, Kumericki:2013br, Berthou:2015oaw, Dupre:2016mai, Moutarde:2018kwr, Moutarde:2019tqa, Kumericki:2016ehc, Dutrieux:2021wll, Almaeen:2024guo, Guo:2025muf, Mezrag:2026wcf}.
It is thus crucial to investigate whether GPDs can {\it in principle} be inferred with full $x$ dependence from a combination of known exclusive observables without relying on any specific parametrization.
This is the main purpose of this paper.

Our choice of processes is guided by a recently proposed framework that unifies all GPD-sensitive reactions into single-diffractive hard exclusive processes (SDHEPs)~\cite{Qiu:2022pla}, which are $2\to3$ processes with a pair of particles carrying large and balancing transverse momenta in the final state, with the diffracted hadron moving in the forward direction.
Varying the external particles of SDHEPs gives all the well-studied exclusive processes, including the above-mentioned ones along with those in~\refs{Brodsky:1994kf, Frankfurt:1995jw, Radyushkin:1996ru, Kumano:2009he, Kumano:2009ky, Berger:2001zn, Goloskokov:2015zsa, Sawada:2016mao, ElBeiyad:2010pji, Pedrak:2017cpp, Grocholski:2021man, Grocholski:2022rqj, Siddikov:2022bku, Siddikov:2023qbd, Boussarie:2016qop, Duplancic:2018bum, Duplancic:2022ffo, Duplancic:2023kwe, Crnkovic:2025man, Qiu:2022bpq, Qiu:2023mrm, Qiu:2024mny, Qiu:2024reu, Nabeebaccus:2026rco}.
Generalizing SDHEPs to allow more than three final-state particles is also straightforward~\cite{Guidal:2002kt, Belitsky:2002tf, Belitsky:2003fj, Deja:2023ahc, Pedrak:2020mfm}.

Because of their exclusive nature, GPDs enter the observables for all these processes at the {\it amplitude} level; the $x$~variable in the GPD is the parton loop momentum fraction and is thus integrated over all possible values from $-1$ to $+1$ in the factorization formula for the amplitude $\M^{[F]}$,
\beq[eq:gpd-integral]
    \M^{[F]}(\xi, t, Q)
    = \int_{-1}^1 \dd{x} C(x, \xi, Q; \mu)\, F(x, \xi, t; \mu),
\eeq
where $C(x, \xi, Q; \mu)$ is the hard scattering coefficient,
with $Q$ an external kinematic variable and $\mu$ the factorization scale.
\footnote{Here we use $Q$ as a generic notation. It can be uniformly chosen as the transverse momentum $q_T$ or polar angle $\theta$ of the two final-state hard particles in any SDHEP~\cite{Qiu:2024reu}. For electroproduction processes, $Q$ is usually chosen to be the virtuality of the exchanged photon.}
The sensitivity of a particular process to the $x$ dependence of GPDs is determined by the form of the hard coefficient $C$, especially the entanglement between $x$ and $Q$ therein.
However, such entanglement is absent at leading order (LO) for DVCS and TCS, which is the origin of the ambiguity emphasized in Ref.~\cite{Bertone:2021yyz}.
Evolution over a range of scales can add new linear combinations of the input GPDs, but does not in general make the deconvolution unique.
Higher-order effects are also usually too small to materially alter this.

To make GPD extraction sensitive to $x$, it is therefore necessary to include processes whose hard coefficients entangle $x$ and $Q$ already at LO.
At the $2 \to 3$ level, two such SDHEPs --- $\gamma\gamma$ mesoproduction and $\gamma\pi$ photoproduction --- have been explored~\cite{Qiu:2022bpq, Qiu:2024mny, Qiu:2022pla, Qiu:2023mrm, Boussarie:2016qop, Duplancic:2018bum, Duplancic:2022ffo, Duplancic:2023kwe} and shown with specific GPD models~\cite{Qiu:2024mny, Qiu:2023mrm} to break the solution degeneracy~\cite{Bertone:2021yyz, Moffat:2023svr}.
In this work, we reverse the logic to study whether a GPD can be uniquely determined with the $x$-$Q$ entangled hard coefficients of these two processes and to what extent its $x$ dependence can be recovered.
Processes beyond $2\to3$ generally also have such entanglement, but their cross sections are strongly suppressed.

Even if a GPD can be uniquely extracted, \eq{eq:gpd-integral} has a further important implication for the quantification of uncertainties.
Suppose a set of measured values of $\M^{[F]}(\xi, t, Q)$ constrains the GPD $F$ down to a unique solution, $\bar{F}$.
The uncertainty $\delta \M$ bounds only the $C$-weighted $x$ integral of the variation 
$\var{F} \equiv F - \bar{F}$, rather than $\var{F}$ point by point. 
The uncertainties of experimental observables will therefore propagate into correlated uncertainties of GPDs at different $x$ points.
\footnote{In real experimental analyses, since the extraction of $x$ dependence relies on the distribution of $Q$, correlations among different bins of $Q$ will also induce correlations in the $x$ space of GPDs. We do not consider this in the present analysis.}
A similar feature applies to transverse momentum dependent (TMD) parton distributions in impact-parameter $b_T$-space: since experimental data are measured in transverse-momentum $k_T$-space and linked to the $b_T$-space distributions via a Bessel transform~\cite{Collins_2011, Boussarie:2023izj}, the data uncertainties likewise propagate into correlated uncertainties across different $b_T$ points.

In contrast, in the global QCD analysis of collinear PDFs, which enter observables at the cross section level, the $x$ dependence can be constrained locally.
Consider, for example, the cross section for inclusive deep-inelastic scattering of a lepton (with four-momentum $\ell$) from a hadron ($p$) to a scattered lepton ($\ell'$) via the exchange of a photon ($q = \ell - \ell'$) with virtuality $Q^2 = -q^2$.
In this case the $x$ variable in the PDF $f(x; \mu)$ is constrained at leading twist by the measured Bjorken variable $x_B$,
\beq[eq:DIS-PDF]
    \frac{\dd\sigma_{\rm DIS}}{\dd{x}_B \dd Q^2}
    = \int_{x_B}^1 \dd{x} \hat{\sigma}\biggpp{ \frac{x_B}{x}, \frac{Q}{\mu} }\, f(x; \mu),
\eeq
where $x_B = Q^2 / 2 p\cdot q$, and we have suppressed the parton flavors as well as other kinematic factors.
Since this integral receives its dominant contribution from $x$ around $x_B$, one may measure the differential cross section at a sufficiently dense grid of $x_B$ such that PDF values at different $x$ points can be constrained independently with local uncertainties.
Their inter-$x$ correlations from sum rule constraints are much weaker than those of GPDs.

To infer the $x$ dependence of GPDs while accounting for inter-$x$ correlations, the analysis must allow GPD values at different $x$ to vary independently.
We choose to represent GPDs by finite elements on a discretized $(x, \xi, t)$ grid, 
as a set of pixels connected by fixed interpolating basis functions,
$F_{ijk} = F(x_i, \xi_j, t_k)$, in a similar fashion to the pixel-based frameworks 
recently developed for GPD evolution~\cite{Freese:2024ypk} and for TMD extraction~\cite{Zaccheddu:2026pyy}.
This representation avoids the bias of parametric models, and, as we will formulate in detail, 
the linearity of the observables in \eq{eq:gpd-integral} with respect to GPDs turns the GPD inversion into an exactly solvable Bayesian inference.
This allows us to quantify not only how well the GPD is determined, but also its uncertainties and inter-pixel correlations.
The observables become linear response matrices on the finite-element basis, which admit a singular value decomposition (SVD) analysis that provides insight into the constraints from various physical observables as well as the polynomiality condition.

As will be demonstrated below, GPDs cannot be inferred point by point in $x$ from currently known observables. 
A large portion of the GPD degrees of freedom remains unconstrained and dominates the pointwise uncertainties.
That said, the well-defined posterior distribution of the GPD pixels still allows us to make predictions for both the experimental observables and other quantities such as the $x$ moments associated with the GPDs themselves.
Three-dimensional imaging of hadrons, which requires the full $x$ dependence, remains a more distant goal.

The rest of this paper is organized as follows.
In \sec{sec:pixelation} we introduce the pixelation setup and develop the finite-element representation.
Next, in \sec{sec:obs} we introduce the physical observables that will be studied for constraining the GPDs.
These observables are all related to GPDs via linear convolutions of the form of \eq{eq:gpd-integral}, and are thus obtained from the GPD pixels by linear transformations.
In \sec{sec:bayesian} we show that the inversion of the GPDs becomes an exactly solvable Bayesian inference with a Gaussian likelihood, and formulate the result in terms of SVD.
After collecting the pseudodata in \sec{sec:data}, we present the detailed results of our analyses in \sec{sec:results}.
Finally, we conclude the paper in \sec{sec:con}.

\section{Pixelated modeling of GPDs}
\label{sec:pixelation}

While the GPDs are functions of $x$, $\xi$ and $t$ (along with the scale $\mu$), in this analysis the variable $t$ enters neither the factorization formula nor the evolution equations, and the GPD reconstruction at different values of $t$ decouple.
For simplicity, therefore, we shall work at a single, fixed value of $t$, and suppress its dependence in the GPD in the rest of the analysis,
$F^f(x, \xi, t; \mu) \to F^f(x, \xi)$, for parton flavors $f = u, d, s$ and $g$.
\footnote{Including the $t$ dependence as an additional grid dimension would be straightforward. It would also allow the constraint from the forward limit $\xi, t \to 0$, in which the GPD reduces to the PDF, provided that certain smoothness conditions are imposed on the pixels.}
For the pixelation construction, we model $F^f(x, \xi)$ by its values $\cc{ F^f(x_i, \xi_j) }$ on a two-dimensional grid of $x$ and~$\xi$ at an initial scale $\mu_0$ for each type of GPD, $F^f = H^f$, $E^f$, etc.
At any other scale $\mu$, the GPD values are obtained from the evolution equation~\cite{Freese:2024ypk}.
The pixel values are treated as independent parameters, with no functional form assumed or smoothness condition imposed among them, which reduces the parametrization bias to as minimal level as possible.

\subsection{Finite element method}
\label{ssec:fem}

For simplicity, we choose {\it linearly} spaced $x$ and $\xi$ grids in the ranges $x \in [-1, 1]$ and $\xi \in [0, 1]$, respectively.
These grids do not need to coincide with each other, although we require that the diagonal GPD values $F^f(\pm\xi, \xi)$ are available on the grid for each $\xi$.
Denoting the nodes of $x$ and $\xi$ by
\beq[eq:x-xi-grid]
	x_i = -1 + i \, \Delta x, \qquad
	\xi_j = j \, \Delta \xi,
\eeq
for $i = 0$, $1$, $\ldots$, $N_x - 1$, and $j = 0$, $1$, $\ldots$, $N_\xi - 1$,
with
$$
	\Delta x
    = \frac{2}{N_x - 1},
    \qquad
    \Delta \xi
    = \frac{1}{N_\xi - 1},
$$
we require $\Delta \xi$ to be an integer multiple of $\Delta x$, and $N_x$ to be odd.

Away from the $x$-$\xi$ nodes, the GPD value is obtained by polynomial interpolation.
Following Freese {\it et al.}~\cite{Freese:2024ypk}, this can be framed as a finite element method, with the interpolated GPD written as
\beq[eq:interp-gpd]
	P\big[\{ F_{ij}^f \}\big](x, \xi)
	= \sum_{i, j} F_{ij}^f \, G^{(p)}_{ij}(x, \xi),
\eeq
where $\cc{ F_{ij}^f \equiv F^f(x_i, \xi_j) }$ is the set of GPD values at the $x$-$\xi$ nodes.
The basis functions $G^{(p)}_{ij}(x, \xi)$, referred to as ``interpixel functions'' (with $p$ denoting the interpolation order), are continuous functions of $x$ and $\xi$ that interpolate between the grid nodes and satisfy
\beq
	G^{(p)}_{ij}(x_{a}, \xi_{b})
	= \delta_{i a} \delta_{j b},
\eeq
so that each $G^{(p)}_{ij}$ equals $1$ at its own node $(x_i, \xi_j)$ and $0$ at all other nodes.
We choose the two-dimensional $G^{(p)}_{ij}(x, \xi)$ to be a direct product of two one-dimensional interpixel functions,
\beq[eq:2D-interpixel-g]
	G^{(p)}_{ij}(x, \xi)
	= X^{(p)}_i(x)\ \Xi^{(p)}_j(\xi),
\eeq
which are continuous functions of $x$ and $\xi$, respectively, satisfying $X^{(p)}_i(x_{a}) = \delta_{ia}$ and $\Xi^{(p)}_j(\xi_{b}) = \delta_{jb}$ at the grid nodes.
Specifically, $X^{(p)}_i(x)$ and $\Xi^{(p)}_j(\xi)$ are constructed from local $p$-th order Lagrange polynomials, as detailed in \refs{Freese:2024ypk, Zaccheddu:2026pyy} (see \refs{Bertone:2013vaa, Salam:2008qg} for a similar approach).

The expression for the interpolated GPD in \eq{eq:interp-gpd} applies to GPDs of any flavor $f$ and at any renormalization or factorization scale $\mu$ (we take the two to be equivalent).
Since the interpixel basis $G^{(p)}_{ij}(x, \xi)$ is fixed for a given $p$, it is the grid values $F_{ij}^f = F_{ij}^f(\mu)$ that carry the $\mu$ dependence.
As discussed in \refcite{Freese:2024ypk}, the evolution of GPDs acts as a linear transformation on these values,
\begin{align}
    \frac{\dd F^f_{ij}(\mu)}{\dd \ln\mu^2}
    = \sum_{f'} \sum_{k,\l} K^{(p)}_{ff',ij, k\l}\bigpp{ \alpha_s(\mu) } \,
        F^{f'}_{k \l}(\mu),
\label{eq:dglap}
\end{align}
where $\alpha_s(\mu)$ is the strong coupling.
Here, $K^{(p)}_{ff',ij, k\l}\bigpp{ \alpha_s(\mu) }$ is the ``discrete'' evolution kernel, given by the integral of the ``continuous'' evolution kernel $V_{ff'}\bigpp{ x, y, \xi; \alpha_s(\mu) }$,
\begin{align}
    K^{(p)}_{ff',ij, k\l}\bigpp{ \alpha_s(\mu) }
    &= \int_{-1}^1 \dd y \,
        V_{ff'}\bigpp{ x_i, y, \xi_j; \alpha_s(\mu) } \,
        G_{k\l}^{(p)}(y, \xi_j)
    \nn\\
    &= \delta_{j\l} \int_{-1}^1 \dd y \,
        V_{ff'}\bigpp{ x_i, y, \xi_j; \alpha_s(\mu) } \,
        X_{k}^{(p)}(y),
\label{eq:dglap-k}
\end{align}
where we have used \eq{eq:2D-interpixel-g} in the second step since $\xi$ does not change under the evolution.
The evolution equation (\ref{eq:dglap}) allows us to evolve the GPDs from $\mu^2$ to $\mu^2 + d\mu^2$.
By iteration, we can build an \textit{evolution matrix} $E(\mu^2, \mu_0^2)$ that takes the GPD grid at $\mu_0^2$ to any other scale $\mu^2$,
\beq[eq:evo-mat]
    F^f_{ij}(\mu)
    = \sum_{f'}\sum_{k} E_{ff', ik, j}(\mu^2, \mu_0^2) \,
        F^{f'}_{kj}(\mu_0^2),
\eeq
where the $\xi$ index $j$ stays unchanged due to \eq{eq:dglap-k} and is not summed over.
The evolution matrix is independent of the GPD values, and can therefore be computed once and stored on disk.

Following \eq{eq:interp-gpd}, any observable that can be factorized into the GPD as in \eq{eq:gpd-integral} can be expressed in pixelated form as
\begin{align}
    \M^{[F]}(\xi, Q)
    &= \int_{-1}^1 \dd{x} P\big[\{ F^f_{ij} \}\big](x, \xi; \mu) \, C(x, \xi, Q; \mu)
    = \sum_{i,j} C^{(p)}_{ij}(\xi, Q; \mu) \, F^f_{ij}(\mu),
\label{eq:obs-mat-prod}
\end{align}
which is simply a matrix product on the GPD grid, with the {\it observable matrix} being
\beq[eq:obs-mat]
    C^{(p)}_{ij}(\xi, Q; \mu)
    = \int_{-1}^1 \dd{x}\, G^{(p)}_{ij}(x, \xi)\, C(x, \xi, Q; \mu).
\eeq
In \eq{eq:obs-mat-prod} both $\xi$ and $Q$ are external kinematic variables.
The value of $\xi$ does not need to coincide with the $\xi$ nodes of the GPD grid, so the sum over $j$ on the right-hand side simply serves as an interpolation.
In contrast, the sum over $i$ plays the role of completely integrating over $x$.
In practice, we use the interpolation order $p = 3$ throughout this paper.
\footnote{We found that using interpolation orders $p = 1$ and $p = 5$ for all the analyses in this paper 
does not appreciably change our results or alter our conclusions.}

\subsection{Double distribution layer}
\label{ssec:dd}

Physically, the GPD pixels are not all independent.
A nontrivial constraint comes from polynomiality~\cite{Ji:1998pc}, which correlates the pixels at different values of $\xi$.
To automatically satisfy the polynomiality condition, we introduce an underlying grid of the double distribution (DD)~\cite{Muller:1994ses, Radyushkin:1997ki, Radyushkin:1998es, Radyushkin:1998bz, Radyushkin:2000uy} $D_F^f(\beta, \alpha)$ and construct the GPD by
\beq[eq:dd]
	F^f(x, \xi) = \int_{-1}^1 \dd\beta \int_{-1 + |\beta|}^{1 - |\beta|} \dd\alpha \, \delta(x - \beta - \xi \alpha) \, D_F^f(\beta, \alpha),
\eeq
for each GPD type $F$ and each parton flavor $f$.
We neglect the possible $D$-term~\cite{Polyakov:1999gs} in this paper; it can be added straightforwardly.
The symmetry $F^f(x, \xi) = F^f(x, -\xi)$ with respect to $\xi$ implies the symmetry
\beq[eq:DD-sym-alpha]
    D_F^f(\beta, \alpha) = D_F^f(\beta, -\alpha)
\eeq
of the DD with respect to $\alpha$.
The $n$-th $x$ moment of the GPD,
\beq[eq:gpd-x-moments]
    F^f_n(\xi) \equiv \int_{-1}^1 \dd{x} x^{n-1} F^f(x, \xi),
\eeq
is then an even polynomial in $\xi$ of degree $n - 1$ or lower.

The $(\beta, \alpha)$ grid of the DD is restricted to the rhombic region $|\alpha| + |\beta| \leq 1$.
This is conveniently constructed by first making a change of variables,
\beq[eq:dd-grid-uv]
	\alpha = \frac{u + v}{\sqrt{2}}, \qquad
	\beta = \frac{u - v}{\sqrt{2}},
\eeq
which is a rotation of axes by $45^\circ$, and then choosing $N_u = N_v$ nodes of $u$ and $v$ in $\big[\!-\!\tfrac{1}{\sqrt2}, \tfrac{1}{\sqrt2}\big]$, both linearly spaced with the same spacing and centered at $0$ such that the points with $\alpha = 0$ or $\beta = 0$ lie exactly on the grid.
Following the same finite element method as in \eq{eq:interp-gpd}, a DD $D$ is then specified by a set of values on the grid $(u_k, v_{\l})$,
\beq
    \cc{
        D_{F, k\l}^{f}
        \equiv D^{f}_{F}\biggpp{ \frac{u_k - v_{\l}}{\sqrt{2}}, \frac{u_k + v_{\l}}{\sqrt{2}}}
    },
\eeq
and its values off the nodes are given by local Lagrange interpolation.
This can be written as a linear combination of the DD interpixels $B^{(p)}_{k\l}(u, v)$,
\beq[eq:interp-dd]
    P\big[\{ D^{f}_{F, k\l} \}\big](\beta, \alpha)
    = \sum_{k, \l} D^{f}_{F, k\l} \,
        B^{(p)}_{k\l}\biggpp{
            \frac{\alpha + \beta}{\sqrt{2}},
            \frac{\alpha - \beta}{\sqrt{2}}
        },
\eeq
where $p$ is the interpolation order, and we again take the two-dimensional $B^{(p)}_{k\l}(u, v)$ to be the direct product of two one-dimensional interpixels of $u$ and $v$,
\beq
    B^{(p)}_{k\l}(u, v)
    = U^{(p)}_k(u) \, U^{(p)}_{\l}(v),
\eeq
with $U^{(p)}$ the same for the $u$ and $v$ dimensions and constructed similarly to $X^{(p)}$ and $\Xi^{(p)}$.

Substituting \eq{eq:interp-dd} into \eq{eq:dd},
we construct the GPD values $\{ F^{f}_{ij} \}$ on the $(x, \xi)$ grid via a linear transformation of the DD values $\{ D^{f}_{F, k\l} \}$,
\beq[eq:dd2gpd]
    F^{f}_{ij}
    = \sum_{k, \l} R^{(p)}_{ij, k\l} \, D^{f}_{F, k \l},
\eeq
where the \textit{Radon matrix} $R^{(p)}_{ij, k\l}$ is given by the Radon transformation of the DD interpixels,
\begin{align}
    R^{(p)}_{ij, k\l}
    &= \int_{-1}^1 \dd\beta \int_{-1 + |\beta|}^{1 - |\beta|} \dd\alpha \, \delta(x_i - \beta - \xi_j \alpha) \,
    B^{(p)}_{k\l}\biggpp{
            \frac{\alpha + \beta}{\sqrt{2}},
            \frac{\alpha - \beta}{\sqrt{2}}
        },
\label{eq:radon-mat}
\end{align}
which can be computed with any suitable quadrature and stored on disk once and for all.
The DDs are then connected to the observables through the product of the Radon matrix, the evolution matrix in \eq{eq:evo-mat}, and the observable matrix in \eq{eq:obs-mat}.

\section{Observables}
\label{sec:obs}

Our aim in this paper is to address whether GPDs can be extracted with their full $x$ dependence from a combination of known exclusive processes, and how well this can be achieved in practice.
For the purpose of this study, we will assume that the basic GPD integrals of the form in \eq{eq:gpd-integral} have been extracted from measured cross sections and azimuthal asymmetries with full flavor separation, arbitrary kinematic coverage, and good precision, and shall refer to them as the ``observables.''
Realizing this assumption is by itself a challenging, if not impossible, task; it defines the idealized setup of our study.

To avoid unnecessary complications that would otherwise obfuscate the essentials of the analysis, we consider observables that depend only on the charge-conjugation-even (C-even) component of the up-quark GPD $H^u$,
\beq
    H^{u+}(x, \xi)
    \equiv H^u(x, \xi) - H^u(-x, \xi),
\eeq
and examine how this can be constrained.
In the pixelation approach, each pixel of this GPD, $H^{u+}_{ij}$ (or the corresponding DD, $D^{u+}_{H, k \l}$, to be further discussed in \sec{sec:bayesian}), is a free parameter.
All other flavors are held fixed, apart from their evolution.
The observables are divided into the following two types.

\subsection{DVCS integral}

The first type of observable is the integral relevant to real-photon electroproduction from a nucleon,
\beq[eq:dvcs]
    e(\l) + N(p)
    \to
    e(\l') + \gamma(q') + N(p'),
\eeq
which contains DVCS as a subprocess.
The hard scale $Q = \sqrt{-(\l - \l')^2}$ is directly associated with the polar angle $\theta$ of the final electron in the SDHEP frame, which universally applies to all SDHEPs~\cite{Qiu:2024reu}, and here is the center-of-mass (c.m.) frame of the final electron and photon, with the initial electron moving along the $-z$ direction.
Explicitly, one has
\beq[eq:Q2-c]
    Q^2 \simeq \frac{\xi \, s}{1 + \xi} (1 + \cos\theta),
\eeq
where $s = (p + \l)^2$ is the overall c.m.\ energy squared, and we have assumed that $t$ and the nucleon mass are negligible relative to $Q^2$.
As pointed out in \refcite{Qiu:2023mrm} and explicitly shown in \refs{Qiu:2024reu, Qiu:2025ksq}, the $\theta$ or $Q$ dependence entirely factors out of the relevant GPD integral at LO,
\beq[eq:dvcs-int]
    \H(\xi; \mu)
    \equiv \int_{-1}^1 \dd x \,
        \frac{H^{u+}(x, \xi; \mu)}{x - \xi + i \epsilon},
\eeq
and in this sense we will refer to this as a {\it scaling} integral.
Note that the $\mu$ dependence of $\H$ on the left-hand side of Eq.~(\ref{eq:dvcs-int}) arises because we retain only the LO hard coefficient.
As a result, even at LO, $Q$ enters the observable through the evolution of the GPD when the factorization scale is set to $\mu = Q$,
\beq
    \H(\xi, Q)
    = \H(\xi; \mu = Q).
\eeq

In terms of \eq{eq:obs-mat}, the scaling integral gives the observable matrix,
\begin{align}
    C^{{\rm SI}\, (p)}_{ij}(\xi)
    &= \int_{-1}^1 \dd{x} G^{(p)}_{ij}(x, \xi) \pp{
        \frac{1}{{x - \xi + i \epsilon}}
        + \frac{1}{{x + \xi - i \epsilon}}
    }
    \nn\\
    &= \Xi^{(p)}_{j}(\xi)
        \bb{ {\rm P.V.}\!\int_{-1}^1 \dd{x} X^{(p)}_{i}(x) \pp{
                \frac{1}{{x - \xi}}
                + \frac{1}{{x + \xi}}
            }
        - i \pi \pp{ X^{(p)}_i(\xi) - X^{(p)}_i(-\xi) }
    }
    \nn\\
    &\equiv \Xi^{(p)}_{j}(\xi)
        \bb{ R^{{\rm SI}\, (p)}_i(\xi)
        + i \, I^{{\rm SI}\, (p)}_i(\xi)
    },
\label{eq:obs-mat-dvcs}
\end{align}
which acts on the GPD pixels $H^{u}_{ij}(\mu)$,
\beq
    \H(\xi; \mu)
    = \sum_{i,j} C^{{\rm SI}\, (p)}_{ij}(\xi) \,
        H^{u}_{ij}(\mu),
\eeq
and contains a real part and an imaginary part,
\begin{subequations}
\begin{align}
    \Re\H(\xi; \mu)
    &= \sum_{i,j} R^{{\rm SI}\, (p)}_i(\xi) \, \Xi^{(p)}_{j}(\xi) \,
        H^{u}_{ij}(\mu),
    \\
    \Im\H(\xi; \mu)
    &= \sum_{i,j} I^{{\rm SI}\, (p)}_i(\xi) \, \Xi^{(p)}_{j}(\xi) \,
        H^{u}_{ij}(\mu).
\label{eq:obs-mat2-dvcs}
\end{align}
\end{subequations}
In the DVCS observable matrices, the interpixel function $\Xi^{(p)}_{j}(\xi)$ is localized around $\xi = \xi_j$, so an observable at a measured $\xi$ constrains only the pixels at the nodes nearest to $\xi$.
The sum over $i$, however, reduces the $x$ dimension to two numbers: the real part, given by a principal value (P.V.) integral, and the imaginary part, which samples the GPD at the ridge $x = \pm\xi$.

Without evolution, or at a fixed scale $\mu_0$, these two constraints are not sufficient to infer the $x$ dependence of the GPD: at least $\tfrac12(N_x - 1) - 2$ of the $\tfrac12(N_x - 1)$ independent pixels remain unconstrained at each $\xi_j$.
This unconstrained subspace of the pixel space is spanned by the shadow GPDs~\cite{Bertone:2021yyz, Moffat:2023svr}.
On the other hand, with evolution included the DVCS observables at different $Q$ give additional constraints,
\begin{align}
    \H(\xi, Q)
    &= \sum_{i,j, k} \sum_f C^{{\rm SI}\, (p)}_{ij}(\xi) \,
        E_{uf, ik, j}(Q^2, \mu_0^2) \, H^{f}_{kj}(\mu_0)
    \nn\\
    &= \sum_{i,j, k} C^{{\rm SI}\, (p)}_{ij}(\xi)
        \biggbb{ E_{uu, ik, j}(Q^2, \mu_0^2) \, H^{u}_{kj}(\mu_0)
            + \sum_{f \neq u} E_{uf, ik, j}(Q^2, \mu_0^2) \, H^{f}_{kj}(\mu_0)
        },
\label{eq:evo-dvcs-mat}
\end{align}
where $\mu_0$ is the initial scale at which the GPD pixels are defined.
The evolution matrix $E$ ``shuffles'' the pixels in the $x$ dimension, by different amounts for different $Q$, before they are contracted with the observable matrix $C$.
This $Q$ dependence induced by the evolution thus gives additional sensitivity to the $x$ dependence.

\subsection{Nonscaling integrals}

The second type of observables consists of the {\it nonscaling} integrals (NSIs) of two $2 \to 3$ SDHEPs.
The first is the mesoproduction of a photon pair with large relative transverse momentum~\cite{Qiu:2022bpq, Qiu:2024mny},
\beq[eq:aa]
    N(p) + \pi(p_2)
    \to
    N'(p') + \gamma(q_1) + \gamma(q_2).
\eeq
Its amplitude contains the GPD integral,
\begin{align}
    I^{\gamma\gamma}(\xi, \theta)
    &= \int_0^1 \dd{z} \frac{\phi(z)}{z(1-z)}
        \bb{ \frac{1 + \cos\theta - 2z}{\sin^2\theta} + \frac{1}{1 + \cos\theta - 2z} }
        \int_{-1}^{1} \dd{x} \frac{H^{u+}(x, \xi; \mu_0)}{x - \xi \, \rho_1(\theta, z) + i \epsilon \, \sgn\!\bb{\cos^2\frac{\theta}{2} - z} },
\label{eq:nsi-meso}
\end{align}
where $\theta$ is the polar angle of one of the final-state photons in the SDHEP frame, $\phi(z)$ is the normalized pion distribution amplitude (DA), and $\rho_1$ is given by
\begin{align}
    \rho_1(\theta, z)
    = \frac{1 - z + \tan^2(\tfrac\theta2) \, z}{1 - z - \tan^2(\tfrac\theta2) \, z}\, ,
\end{align}
which lies in $[1, \infty) \cup (-\infty, -1]$ as $\theta$ or $z$ varies.
This integral contains both real and imaginary parts, with the imaginary part coming exclusively from the DGLAP regions of the GPD.
For a given DA, the $z$ integral can be carried out first to convert \eq{eq:nsi-meso} to
\begin{align}
    I^{\gamma\gamma}(\xi, \theta)
    &= \int_{-1}^{1} \dd{x}
        C_{\phi}^{\gamma\gamma}(x, \xi, \theta) \,
        H^{u+}(x, \xi; \mu_0).
\label{eq:nsi-meso-x}
\end{align}
In either \eq{eq:nsi-meso} or \eqref{eq:nsi-meso-x}, the $\theta$ dependence cannot be separated from the $x$ integral, and hence we call $I^{\gamma\gamma}$ an NSI.
For the special case with the DA taking the asymptotic form $\phi_0(z) = 6z(1-z)$, which we adopt in this paper, the kernel $C_{\phi_0}^{\gamma\gamma}$ is explicitly given in \refcite{Qiu:2024mny}.

As a kinematic crossing of \eq{eq:aa}, the other process we consider is the photoproduction of a photon-meson pair~\cite{Qiu:2022pla, Qiu:2023mrm},
\beq[eq:ap]
    N(p) + \gamma(p_2)
    \to
    N'(p') + \pi(q_1) + \gamma(q_2),
\eeq
which gives a different NSI,
\begin{align}
	I^{\gamma\pi}(\xi, \theta)
	&= \int_0^1 \dd{z} \frac{\phi(z)}{2z(1-z)}
			\bb{ \frac{1}{z + \cos^2(\tfrac\theta2) (1-z)} - \frac{z + \cos^2(\tfrac\theta2) (1-z)}{\cos^2(\tfrac\theta2)} }
			\int_{-1}^{1} \dd{x} \frac{ H^{u+}(x, \xi; \mu_0) }{x - \xi \, \rho_2(\theta, z) + i \epsilon} .
\label{eq:nsi-photo}
\end{align}
Here $\rho_2$ given by
\begin{align}
	\rho_2(\theta, z)
	= \frac{\cos^2(\tfrac\theta2)\, (1-z) - z}{\cos^2(\tfrac\theta2)\, (1-z) + z},
\end{align}
and lies in $[-1, 1]$ as $\theta$ or $z$ varies.
It also has a real part and an imaginary part, but the latter now comes purely from the ERBL region of the GPD.
Similarly to \eq{eq:nsi-meso-x}, the $z$ integral can be carried out to give
\begin{align}
    I^{\gamma\pi}(\xi, \theta)
    &= \int_{-1}^{1} \dd{x}
        C_{\phi}^{\gamma\pi}(x, \xi, \theta) \,
        H^{u+}(x, \xi; \mu_0),
\label{eq:nsi-photo-x}
\end{align}
where, for the asymptotic DA, the kernel is
\begin{align}
    C_{\phi_0}^{\gamma\pi}(x, \xi, \theta)
    &= \bb{1 - \frac{8\xi^2 (1 + c)}{A^2}} \cc{
        \frac{6L}{A}
        + \frac{3(1 - c)}{2(1 + c) \xi} \bb{
            1 +
            \frac{(3 + c) A}{8(1 + c) \xi}
        }
    } - \frac{3(1 - c) \bigbb{
        (1 - c)(3 + c) x
        + (c^2 + 14 c + 17) \xi
    } }{16(1 + c)^2 \xi^2} ,
\end{align}
with $c \equiv \cos\theta$ and
\begin{subequations}
\begin{align}
    A &= 2(x + \xi) - (1 + c)(x - \xi),
    \\
    L &= \ln\pp{ \frac{2}{1 + c} \left|
        \frac{x + \xi}{x - \xi}
        \right|
    } - i \pi \, \theta(\xi^2 - x^2).
\end{align}
\end{subequations}

In both \eqs{eq:nsi-meso-x}{eq:nsi-photo-x}, the scale is fixed at $\mu_0$, and the observable matrices are
\begin{align}
    C^{a}_{ij}(\xi, \theta)
    &= \int_{-1}^{1} \dd{x} G^{(p)}_{ij}(x, \xi) \bigbb{
        C_{\phi_0}^a(x, \xi, \theta)
        - C_{\phi_0}^a(-x, \xi, \theta)
    }
    \nn\\
    &= \Xi^{(p)}_j(\xi) \int_{-1}^{1} \dd{x} X^{(p)}_{i}(x) \bigbb{
        C_{\phi_0}^a(x, \xi, \theta)
        - C_{\phi_0}^a(-x, \xi, \theta)
    },
\label{eq:nsi-mat}
\end{align}
for $a = \gamma\gamma$ or $\gamma\pi$.
These map the GPD grid to the NSI observables,
\beq[eq:nsi-prod]
    I^a(\xi, \theta)
    = \sum_{i, j} C^{a}_{ij}(\xi, \theta) \, H^u_{ij}(\mu_0),
    \qquad
    a = \gamma\gamma, \, \gamma\pi.
\eeq
Similarly to the $Q$ dependence in \eq{eq:evo-dvcs-mat}, $\theta$ introduces extra sensitivity to the $x$ dependence.
Observables at different $\theta$ values provide independent constraints on the GPD pixels.
The difference, however, is that \eq{eq:evo-dvcs-mat} depends on $Q$ (or $\theta$ by \eq{eq:Q2-c}) logarithmically, whereas in \eq{eq:nsi-mat} $\theta$ is entangled with $x$ more strongly through rational functions.

\section{Bayesian Inference as an Exact Gaussian Problem}
\label{sec:bayesian}

In addition to providing a sufficiently flexible representation, the pixelation construction has the advantage of preserving the linearity of the observables with respect to GPDs, as we have explicitly established in \sec{sec:obs}.
This casts the likelihood function of the inversion problem into a Gaussian which can be solved exactly.

\subsection{Bayesian inversion at the GPD level}
\label{ssec:bayes-gpd}

In this work we study the inversion both at the GPD level and at the DD level.
The former quantifies the constraining power of the data alone, while the latter additionally includes the constraint from polynomiality.
We begin first with the inversion at the GPD level, and later extend it to the DD level in \sec{ssec:bayes-dd}.

At the scale $\mu_0$, we retain all the pixels $H^{u+}_{ij}$ of the C-even $u$-quark GPD as free parameters, but fix all other flavors.
These parameters can be collectively denoted by a vector $\bm{\epsilon}$ in the $n_G$-dimensional vector space, 
with $n_G = \tfrac12 N_{\xi} (N_x - 1)$,
\footnote{Due to the C-even combination, only $\tfrac12(N_x - 1)$ out of the $N_x$ pixels per $\xi$ are independent.}
$$
    \bm{\epsilon}
    \equiv (\epsilon_m)
    = \{ H^{u+}_{ij} \},
$$
with $m$ labeling the GPD pixels.
The observables in \sec{sec:obs} are obtained via linear transformations of $\bm{\epsilon}$,
\beq[eq:obs-linear-prod]
    \bm{T}(\bm{\epsilon})
    \equiv \pp{ T_a (\bm{\epsilon})}
    = \bm{A} \cdot \bm{\epsilon} + \bm{B}
    \equiv (A_{a m} \epsilon_m) + (B_a),
\eeq
where sums over repeated indices are understood.
Here, $A_{a m}$ collects the observable matrices ($C$'s) in Eqs.~\eqref{eq:obs-mat-dvcs} and \eqref{eq:nsi-mat}.
The index $a$ labels the data points, each specified in general by the values of $(\xi, Q, \theta)$, and the data set can be any combination of the observables, as specified in \sec{sec:data} below.
For the evolved DVCS integral in \eq{eq:evo-dvcs-mat}, $A_{a m}$ also contains the flavor-diagonal evolution kernel $E_{uu, ik, j}(Q^2, \mu_0^2)$.
Since we include evolution effects only in this observable, other flavors contribute to the left-hand side of \eq{eq:obs-linear-prod} only through the second term of \eq{eq:evo-dvcs-mat}.
They enter entirely into the inhomogeneous term $B_a$ in \eq{eq:obs-linear-prod} and are thus fixed in our setup.

With data $\bm w \equiv (w_a)$ measured for the observables $\bm{T}$, the posterior probability distribution of the pixel parameters $\bm{\epsilon}$ is given by Bayes' theorem,
\begin{align}
    L(\bm{\epsilon})
    = P(\bm{\epsilon} \mid \bm w)
    \propto P(\bm w \mid \bm{\epsilon}) \, P(\bm{\epsilon}),
\end{align}
where $P(a \mid b)$ denotes the conditional probability and an $\bm{\epsilon}$-independent normalization constant has been suppressed.
The last factor $P(\bm{\epsilon})$ is the prior probability distribution of the GPD pixels, which we take as a Gaussian $\mathcal{N}(\bm h_0, \bm V_0)$ with mean $\bm h_0$ and covariance matrix $\bm{V}_0 \equiv ((V_0)_{mm'})$.
Assuming that $P(\bm w \mid \bm{\epsilon})$ is also Gaussian, with covariance $\bm{\Sigma} = (\Sigma_{ab})$, we have the posterior distribution,
\begin{align}
    L(\bm{\epsilon})
    \propto \exp\cc{
        - \frac{1}{2} \bb{ \,
        			\pp{ \bm{w} - \bm{B} - \bm{A} \bm{\epsilon} }^\T \,
				\bm{\Sigma}^{-1}
				\pp{ \bm{w} - \bm{B} - \bm{A} \bm{\epsilon} }
			+ (\bm \epsilon - \bm h_0)^\T \, 
				\bm{V}_0^{-1} (\bm \epsilon - \bm h_0)
        \, }
    }.
\label{eq:likelihood-g}
\end{align}
Clearly, this is also a Gaussian distribution of $\bm{\epsilon}$,
\begin{align}
    L(\bm{\epsilon})
    \propto \exp\cc{
        - \frac{1}{2} \pp{ \bm{\epsilon} - \bm{h}}^\T
        \bm{V}_{\rm post}^{-1}
        \pp{ \bm{\epsilon} - \bm{h}}
    },
\label{eq:post-L-g}
\end{align}
with the inverse covariance matrix,
\beq[eq:posterior-cov-g]
    \bm{V}_{\rm post}^{-1}
    = \bm{A}^\T \bm{\Sigma}^{-1} \bm{A}
        + \bm{V}_0^{-1},
\eeq
and mean,
\beq[eq:posterior-mean-g]
    \bm{h}
    = \bm{V}_{\rm post} \bb{
        \bm{A}^\T \bm{\Sigma}^{-1}
        \pp{ \bm{w} - \bm{B} }
        + \bm{V}_0^{-1} \, \bm{h}_0
    }.
\eeq

\subsection{Insights from singular value decomposition analysis}
\label{ssec:svd-g}

We note that without the prior term $\bm{V}_0^{-1}$, the right-hand side of \eq{eq:posterior-cov-g} may not be invertible.
This is because the observable matrix $\bm{A}$ generally reduces the information encoded in the GPD when converting it to observables.
For example, if we only have the DVCS observables in \eq{eq:obs-mat2-dvcs} at the fixed scale $\mu_0$, the dimension of $\bm{w}$ is just $2$ times the number of measured $\xi$ values, so that the rank of the corresponding observable matrix $\bm{A}$ is at most 2 per~$\xi$.
With the prior term as a regulator, $\bm{V}_{\rm post}^{-1}$ is bounded from below by $\bm{V}_0^{-1}$ and becomes positive-definite and thus invertible.

More insight into the inversion is gained by formulating the result in terms of singular value decomposition (SVD).
To this end, it is convenient to renormalize the pixel vector as $\bm\epsilon = \sqrt{\bm V_0} \, \bm z$.
The vector $\bm z$ has the same dimension as $\bm\epsilon$, with a prior normal distribution centered at $\bm z_0 = \bm V_0^{-1/2} \bm h_0$ and a posterior Gaussian that has an inverse covariance matrix,
\beq[eq:cov-z-g-inv]
	\bm V_z^{-1} 
		= \sqrt{\bm V_0} \, \bm{V}_{\rm post}^{-1} \, \sqrt{\bm V_0} 
		= \mathbf{1} + \sqrt{\bm V_0} \, \bm{A}^\T \bm{\Sigma}^{-1} \bm{A} \, \sqrt{\bm V_0},
\eeq
and a mean,
\beq[eq:mean-z-g-inv]
	\bm \mu_z
		= \bm V_z \bb{
			\sqrt{\bm V_0} \, \bm{A}^\T \bm{\Sigma}^{-1} (\bm w - \bm B)
			+ \bm z_0
		}.
\eeq
Both $\bm V_0$ and $\bm \Sigma$ here are positive definite, so have well defined matrix square roots.

Next, we define the observable matrix normalized to both the prior and the data uncertainties,
\beq[eq:J-def]
	\bm J = \bm \Sigma^{-1/2} \, \bm{A} \, \sqrt{\bm V_0},
\eeq
and carry out its SVD,
\beq[eq:J-svd]
	\bm J = \sum_{k = 1}^{r_A} \lambda_k \, \bm u^{(k)} \, \bm v^{(k)\T},
\eeq
where $r_A$ is the rank of $\bm A$, which equals that of $\bm J$, since $\bm\Sigma^{-1/2}$ and $\sqrt{\bm V_0}$ are invertible, and $\lambda_1 \geq \lambda_2 \geq \cdots \geq \lambda_{r_A} > 0$ are the singular values of $\bm J$.
The $\bigcc{ \bm u^{(k)} \equiv (u^{(k)}_a) }_{k = 1}^{r_A}$ and $\bigcc{ \bm v^{(k)} \equiv (v^{(k)}_m) }_{k = 1}^{r_A}$ are two sets of orthonormal vectors in the observable space and in the GPD space, respectively.
Equation~(\ref{eq:cov-z-g-inv}) then becomes
\begin{align}
	\bm V_z^{-1} 
		= \mathbf{1} + \bm J^\T \bm J
		= \sum_{k = 1}^{n_G} \pp{ 1 + \lambda_k^2 } \bm v^{(k)} \, \bm v^{(k)\T},
\label{eq:cov-z-g-inv-svd}
\end{align}
and thus
\begin{align}
	\bm V_z
		= (\mathbf{1} + \bm J^\T \bm J)^{-1}
		= \sum_{k = 1}^{n_G} \frac{\bm v^{(k)} \, \bm v^{(k)\T}}{ 1 + \lambda_k^2 }.
\label{eq:cov-z-g-svd}
\end{align}
Here, we have defined $\lambda_k = 0$ if $k > r_A$ for notational simplicity, and have completed the orthonormal basis in the GPD space to $\bigcc{ \bm v^{(k)} }_{k = 1}^{n_G}$.
The expression in \eq{eq:cov-z-g-svd} indicates that the width of the $\bm z$ distribution along the $\bm v^{(k)}$ direction is reduced from 1 in the prior to $1 / \sqrt{ 1 + \lambda_k^2 }$ in the posterior.
Only the directions with $\lambda_k \gg 1$ are well constrained, hence we can define the effective number of constraints provided by the given observable set as
\beq[eq:kappa]
	\kappa = \sum_{k = 1}^{n_G} \pp{ 1 - \frac{1}{1 + \lambda_k^2} }
		= \tr\pp{ \mathbf{1} - \bm V_z },
\eeq
where ``$\tr$'' takes the matrix trace in GPD space.

Applying \eqs{eq:J-svd}{eq:cov-z-g-svd} to \eq{eq:mean-z-g-inv}, we obtain the posterior mean of the GPD vector in terms of $\bm z$,
\begin{align}
	\bm \mu_z
	&= \sum_{k = 1}^{n_G} \frac{\bm v^{(k)} \, \bm v^{(k)\T}}{ 1 + \lambda_k^2 } 
		\biggcc{ \sum_{l = 1}^{r_A} \lambda_l \, \bm v^{(l)}
			\bb{ \bm u^{(l)\T}
				\bm{\Sigma}^{-1/2} (\bm w - \bm B)
			}
			+ \bm z_0
		}
	\nn\\
	&= \sum_{k = 1}^{r_A} \frac{\lambda_k}{ 1 + \lambda_k^2 } \bm v^{(k)}
			\bb{ \bm u^{(k)\T}
				\bm{\Sigma}^{-1/2} (\bm w - \bm B)
			}
		+ \sum_{k = 1}^{n_G} \frac{\bm v^{(k)}}{ 1 + \lambda_k^2 } \bb{ \bm v^{(k)\T} \bm z_0 },
\label{eq:post-mean-z-svd}
\end{align}
which separates the impacts of the data and the prior mean.
In this paper, the central values of the data are given by the prediction of a chosen GPD model, $\bar{\bm\epsilon} = \sqrt{\bm V_0}\, \bar{\bm z}$, so that $\bm w$ can be written as
\beq[eq:data-fluc]
	\bm w = \overline{\bm w} + \sqrt{\bm \Sigma} \, \delta {\bm w}
	= \bm B + \bm A \sqrt{\bm V_0} \bar{\bm z} + \sqrt{\bm \Sigma} \, \delta {\bm w},
	\qquad
	\delta {\bm w} \sim \N(0, \mathbf{1}),
\eeq
with $\delta {\bm w}$ the normalized noise.
Substituting this into \eq{eq:post-mean-z-svd} further gives
\begin{align}
	\bm \mu_z
	&= \sum_{k = 1}^{r_A} \frac{\lambda_k}{ 1 + \lambda_k^2 } \bm v^{(k)}
			\bb{ \bm u^{(k)\T}
				(\bm J \bar{\bm z} + \delta {\bm w})
			}
		+ \sum_{k = 1}^{n_G} \frac{\bm v^{(k)}}{ 1 + \lambda_k^2 } \bb{ \bm v^{(k)\T} \bm z_0 }
	\nn\\
	&= \sum_{k = 1}^{r_A} \frac{\lambda_k^2}{ 1 + \lambda_k^2 } \bm v^{(k)}
			\bb{ \bm v^{(k)\T} \bar{\bm z} }
		+ \sum_{k = 1}^{r_A} \frac{\lambda_k}{ 1 + \lambda_k^2 } \bm v^{(k)}
			\bb{ \bm u^{(k)\T} \delta {\bm w} }
		+ \sum_{k = 1}^{n_G} \frac{1}{ 1 + \lambda_k^2 } \bm v^{(k)} \bb{ \bm v^{(k)\T} \bm z_0 }.
\label{eq:post-mean-z-svd-2}
\end{align}
This expression shows the different roles of the ``ground-truth'' GPD in the first term, the data noise in the second term, and the prior mean in the last term:
\begin{itemize}
\item the ``truth'' GPD $\bar{\bm z}$ is recovered only along the directions with $\lambda_k \gg 1$ and is suppressed otherwise;
\item the data noise $\delta {\bm w}$ affects the directions with $\lambda_k \sim 1$, contributing an error $\lambda_k / (1 + \lambda_k^2)$ to the $\bm v^{(k)}$ direction;
\item the prior mean $\bm z_0$ survives along the directions with $\lambda_k \ll 1$.
\end{itemize}

Qualitatively, we regard those directions with $0.1 < \lambda_k^2 < 10$ as being sensitive to the data noise in the second term of \eq{eq:post-mean-z-svd-2}.
Compared with the standard deviation $1 / \sqrt{1 + \lambda_k^2}$ given by \eq{eq:cov-z-g-svd}, the shift of the mean caused by the data noise is always smaller,
\beq[eq:noise-ratio]
	\frac{\lambda_k}{ 1 + \lambda_k^2 } 
	\bigg/
	\frac{1}{ \sqrt{1 + \lambda_k^2} } 
	= \frac{\lambda_k}{ \sqrt{1 + \lambda_k^2} } 
	< 1,
\eeq
which, however, approaches $1$ for $\lambda_k \gg 1$.
The directions with $\lambda_k^2 \geq 10$ are well determined, along which the mean reproduces the ground-truth GPD, being minimally affected by the data noise and constrained within the small standard deviation of order $1 / \lambda_k$.
In contrast, the directions with $\lambda_k^2 \leq 0.1$ are weakly constrained, with both means and standard deviations dominated by the prior.

The directions along the right singular vectors $\bm v^{(k)}$ with $\lambda_k^2 \leq 0.01$ are effectively unconstrained.
We define the subspace they span as the shadow GPD subspace,
\beq[eq:shadow-space]
	\S_z = \operatorname{span}\cc{ \bm v^{(k)}: \, 
		\lambda_k^2 \leq 0.01
	},
\eeq
along which the posterior width is reduced by less than $0.5\%$ from the prior width, the posterior mean retains more than $99\%$ of the prior mean, and the noise is at most $0.1$ in units of the prior width.
The directions with $0.01 < \lambda_k^2 \leq 0.1$ lie between the shadow and the moderately constrained ones; they are still dominated by the prior but not invisible to the data, so in the following we will consider them in the constrained subspace.
Labeling the $\bm v^{(k)}$ in \eq{eq:shadow-space} as $\cc{ \bm v^{(n_{\ob} + 1) }, \bm v^{(n_{\ob} + 2) } \cdots, \bm v^{(n_G)} }$, the remaining $\bm v^{(k)}$'s span the ``observed'' GPD subspace of dimension $n_{\ob}$,
\beq
	\O_z = \operatorname{span}\cc{ \bm v^{(1)}, \, 
		\bm v^{(2)}, \, 
		\cdots, \,
		\bm v^{(n_\ob)}
	},
\eeq
to which the data are sensitive. (Note that we necessarily have $n_{\ob} \leq r_A$.)
We can then define the projection of any GPD vector onto the observed subspace,
\beq[eq:P-ob]
	\widehat P_\ob = \sum_{k = 1}^{n_\ob} \bm v^{(k)} \bm v^{(k) \T},
\eeq
and the orthogonal projection onto the shadow subspace,
\beq[eq:P-sh]
	\widehat P_\sh = \mathbf{1} - \widehat P_\ob = \sum_{k = n_\ob + 1}^{n_G} \bm v^{(k)} \bm v^{(k) \T},
\eeq
with
\beq[eq:P-orthog]
	\widehat P_\sh \widehat P_\ob = \widehat P_\ob \widehat P_\sh = 0.
\eeq

When proceeding from the above SVD analysis back to the $x$-$\xi$ pixel space, which amounts to specifying the components $v^{(k)}_m$, the covariance of GPD pixels $\epsilon_m$ is given by
\beq[eq:cov-g]
    \vv{(\epsilon_m - h_m)(\epsilon_{m'} - h_{m'})}
    = \pp{ V_{\rm post} }_{mm'}
    = \sum_{\overline{m}, \overline{m}'} (\sqrt{V_0})_{m\overline{m}} (\sqrt{V_0})_{m'\overline{m}'} (V_z)_{\overline{m} \overline{m}'}\, .
\eeq
We will choose the prior covariance $V_0$ to be proportional to the identity matrix, in which case the pixel-wise distribution width $\sigma_m$ is given by
\beq[eq:sigma-m]
	\sigma_m^2 
		= \vv{(\epsilon_m - h_m)(\epsilon_{m} - h_{m})}
		= (V_0)_{mm} (V_z)_{mm}\, ,
\eeq
and the inter-pixel correlation is 
\beq[eq:cor-g]
    \rho_{mm'}
    = \frac{ \pp{ V_{\rm post} }_{mm'} }{\sqrt{\pp{ V_{\rm post} }_{mm} \pp{ V_{\rm post} }_{m'm'} }}
    = \frac{ (V_z)_{mm'} }{ \sqrt{ (V_z)_{mm} (V_z)_{m'm'} } },
\eeq
where the repeated indices here are not summed.
By rewriting \eq{eq:cov-z-g-svd} as 
\begin{align}
	\bm V_z
		= \mathbf{1} - \sum_{k = 1}^{n_G} \frac{\lambda_k^2}{ 1 + \lambda_k^2 } \bm v^{(k)} \, \bm v^{(k)\T},
\label{eq:cov-z-g-svd-2}
\end{align}
we can see that its diagonal elements, and hence the $\sigma_m$ of individual pixels, are dominated by the less constrained singular vectors, which include those with $0.1 < \lambda_k^2 < 10$ that are susceptible to the data noise, and those with $\lambda_k^2 \leq 0.1$ that include the shadow GPDs with $\lambda_k^2 \leq 0.01$.
The off-diagonal elements, which give the inter-pixel correlations, receive contributions essentially only from the constrained singular vectors, since the shadow GPDs enter with weights $\lambda_k^2 / (1 + \lambda_k^2) \leq 0.01$.
We will see in \sec{sec:results} below that for most sets of observables, the best constrained singular vectors $\bm v^{(k)}$ are localized around the ridge $|x| = \xi$, and thus the correlation generally concentrates around the ridge with short correlation length.

\subsection{Bayesian inversion at the DD level}
\label{ssec:bayes-dd}

The inversion can be performed also at the DD level, by treating the pixels of $\{ (D^{u+}_{H})_{k\l} \}$ as the free parameters at $\mu_0$ and fixing all others.
Denoting them as a vector $\bm{\epsilon}'$, the observables $\bm{T}$ in \eq{eq:obs-linear-prod} are obtained by substituting $\bm{\epsilon} \to \bm{R} \bm{\epsilon}'$, where $\bm{R}$ is the matrix notation of the Radon matrix $R^{(p)}_{ij, k\l}$ in \eq{eq:dd2gpd}.
The same analysis as above can then be carried out to give the posterior probability distribution of $\bm{\epsilon}'$,
\begin{align}
    L_{\rm DD}(\bm{\epsilon}')
    \propto \exp\cc{
        - \frac{1}{2} \pp{ \bm{\epsilon}' - \bm{d}}^\T
        \pp{ \bm{V}_{\rm post}^{\prime} }^{-1}
        \pp{ \bm{\epsilon}' - \bm{d}}
    },
\label{eq:post-L-d}
\end{align}
which is also a Gaussian, with the inverse covariance matrix,
\beq[eq:posterior-cov-d]
    \pp{ \bm{V}_{\rm post}' }^{-1}
    = \bm{R}^\T \bm{A}^\T \bm{\Sigma}^{-1} \bm{A} \, \bm{R}
        + (\bm{V}_0')^{-1},
\eeq
and mean,
\beq[eq:posterior-mean-d]
    \bm{d}
    = \bm{V}'_{\rm post} \bb{
        \bm{R}^\T \bm{A}^\T \bm{\Sigma}^{-1}
        \pp{ \bm{w} - \bm{B} }
        + (\bm{V}_0')^{-1} \, \bm{d}_0
    },
\eeq
where we have taken the prior of $\bm{\epsilon}'$ as a Gaussian with mean $\bm{d}_0$ and covariance matrix $\bm{V}_0'$.
Once $L_{\rm DD}(\bm{\epsilon}')$ is obtained, the probability distribution of 
the GPD vector $\bm{\epsilon}$ can be obtained by the mapping $\bm{\epsilon} = \bm{R} \bm{\epsilon}'$, which still gives a Gaussian, with the mean $\bm{R} \bm{d}$ and covariance matrix $\bm{R} \bm{V}_{\rm post}' \bm{R}^\T$.

An SVD analysis similar to that in \sec{ssec:svd-g} can be carried out for the DD pixels.
By renormalizing $\bm\epsilon' = \sqrt{\bm V_0'} \, \bm z'$, all the results there hold, with $\bm J$ replaced by
\beq[eq:J-dd]
	\bm J' = \bm \Sigma^{-1/2} \, \bm A \, \bm R \, \sqrt{\bm V_0'},
\eeq
and $n_G$ replaced by the number of independent DD pixels, $n_D$.
In particular, the effective number of constraints at the DD level is given by \eq{eq:kappa} with the singular values of $\bm J'$, and the shadow GPDs are the images $\bm R \sqrt{\bm V_0'} \, \bm v'^{(k)}$ of the right singular vectors $\bm v'^{(k)}$ of $\bm J'$ with small singular values.
Compared with the GPD-level inversion, the role of $\sqrt{\bm V_0}$ is now played by the $n_G \times n_D$ matrix $\bm S \equiv \bm R \sqrt{\bm V_0'}$, which induces the prior covariance $\bm R \bm V_0' \bm R^\T = \bm S \bm S^\T$ of the GPD pixels.
The SVD of $\bm S$ does not involve the data and characterizes the polynomiality condition imposed through the DD: the GPD directions associated with small singular values of $\bm S$ have narrow prior widths, so they are suppressed before any data are included.
Therefore, the SVD of $\bm S$ quantifies the constraint from polynomiality, and that of $\bm J'$ quantifies the constraint from the data on top of it.
In the GPD space, the posterior mean is $\bm S \bm\mu_{z'}$ and the covariance is $\bm S \bm V_{z'} \bm S^\T$, where $\bm\mu_{z'}$ and $\bm V_{z'}$ are given by \eqs{eq:post-mean-z-svd-2}{eq:cov-z-g-svd}, respectively, with the SVD of $\bm J'$. 
We will study both SVDs numerically in \sec{sec:results-dd} below.

In deciding how the DD grid should be chosen relative to the GPD grid, we note that besides considerations from numerical accuracy, different DD resolutions will lead to different degrees of correlation among the GPD pixels.
Theoretically, it is known~\cite{Radyushkin:1998bz, Teryaev:2001qm, Belitsky:2000vk} that the Radon transformation $\bm{R}$ from DD to GPD is not invertible, which means that the DD should contain more information than the GPD.
In the pixelation construction, the number of independent DD pixels on each $(u, v)$ plane is of order $\tfrac14 N_u^2$ (for $N_u \gg 1$), where the $\tfrac14$ factor is due to the symmetry in \eq{eq:DD-sym-alpha} and the C-even property, and the number of independent GPD pixels on each $(x, \xi)$ plane is of order $\tfrac12 N_x N_{\xi}$, so that we need at least $\frac14 N_u^2 \geq \tfrac12 N_x N_{\xi}$.
Varying around this, we will consider three benchmarks in this analysis,
\beq[eq:info-ratio]
    \frac{N_u^2}{2 N_x N_{\xi}}
    = \frac{1}{2}, \; 1, \; \mbox{or} \,\; 2,
\eeq
which we will refer to as the half-, equal-, and double-information scenarios, respectively.
We have checked numerically that the GPDs obtained from the DDs in these scenarios satisfy the polynomiality condition, whereas for a larger ratio $N_u^2 / (2 N_x N_{\xi})$, when the DD pixels become much finer than the GPD pixels, polynomiality is no longer maintained on the GPD grid. \\

\section{Pseudodata generation}
\label{sec:data}

\begin{figure}[b]
    \centering
    \includegraphics[scale=0.95]{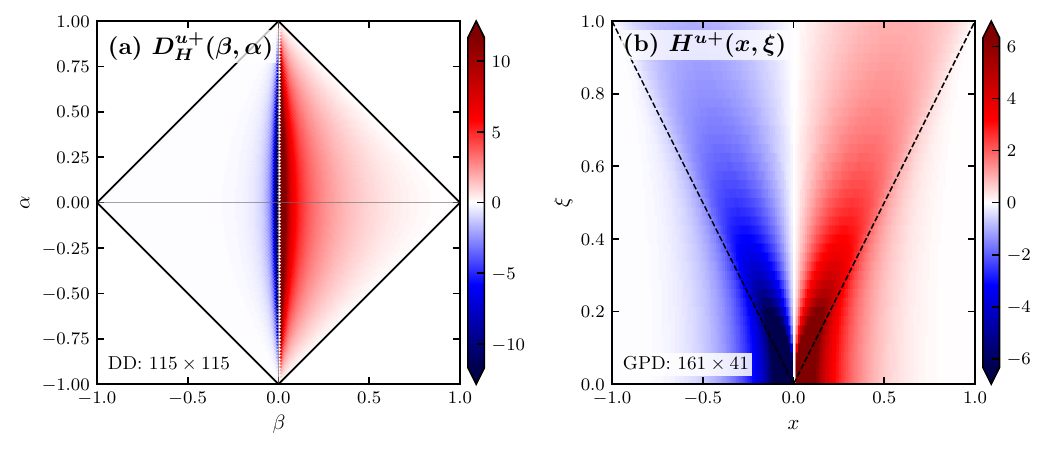}
    \caption{Pixel representations of the ground-truth (a) DD $D_H^{u^+}(\beta,\alpha)$ with $(N_u,N_v)=(115,115)$ and (b) GPD $H^{u^+}(x,\xi)$ with $(N_x,N_\xi)=(161,41)$ from the GK model, for the C-even $u$-quark component, at $\mu_0 = 2~\GeV$. The black solid boundary in (a) marks $|\beta|+|\alpha|=1$, and the dashed lines in (b) mark the ridges $x=\pm\xi$ separating the DGLAP and ERBL regions.}
\label{fig:gpd-image-gk}
\end{figure}

In the numerical analysis that will follow, we fix the GPD grid to $(N_x, N_\xi) = (161, 41)$ and use the Goloskokov--Kroll (GK) model~\cite{Goloskokov:2005sd, Goloskokov:2007nt, Goloskokov:2009ia, Kroll:2012sm} as the ``ground truth'' at $\mu_0 = 2~\GeV$.
\footnote{We do not include the evolution modeled by GK, which corresponds to setting $Q = Q_0$ in their model, 
and since we work at a single $t$ we use their parametrization at $t = 0$.}
The three DD grids in \eq{eq:info-ratio} are correspondingly chosen as $N_u = N_v = 81$, $115$, or $163$.
The images of the GPD and the equal-information DD are displayed in \fig{fig:gpd-image-gk}.

For the GPD and DD of the GK model we use the symbols 
$\widebar{H}^{u+}$ (or $\bar \epsilon$) and $\widebar{D}^{u+}_H$ (or $\bar \epsilon'$), respectively, 
in order to distinguish these from the free pixel parameters.
Following \mbox{\eq{eq:data-fluc}}, 
the pseudodata $w_a$ for the observables in \mbox{\sec{sec:obs}} are generated 
as Gaussian fluctuations around the theoretical expectation values $\bar w_a$,
which are obtained by contracting the GPD pixels with the observable matrices, 
$\bar w_a = T_a(\bar{\bm{\epsilon}}) = A_{am} \bar \epsilon_m + B_a$,
and we assign to them uncorrelated uncertainties,
\beq[eq:data-sigma]
	\Sigma_{ab} = r^2 \bar w_a^2 \, \delta_{ab}\,,
\eeq
with a uniform relative uncertainty $r$ across all the data points.
\footnote{In our numerical treatment we discard data points lying at an isolated zero crossing, 
$|\bar w_a| < 0.1 \cdot \max(|\bar w_{a-1}|, |\bar w_{a+1}|)$, 
along the continuous scan of the data dimension (which is $\xi$ for the DVCS integrals and $\cos\theta$ for the NSIs).}
We take a small $r = 0.01$ as our canonical choice to probe the ``theoretical'' constraining power of each observable, but vary it to $0.1$ and $0.2$ to mimic realistic experimental uncertainties in \sec{sec:results} below.

For the GPD pixel vector $\bm \epsilon$, we take the Gaussian prior in \eq{eq:likelihood-g} with a uniform standard deviation, $\bm V_0 = \eta_H^2 \mathbf{1}$, where $\eta_H = 10$ is much larger than the GPD values, and a vanishing prior mean, $\bm h_0 = \bm 0$.
For each set of the observables in \sec{sec:obs}, we choose the number of data points to saturate the corresponding effective number of constraints $\kappa$ in \eq{eq:kappa}.
Since we assign a fixed relative uncertainty to each data point, adding data points always increases $\kappa$.
However, as the constraining power saturates, an additional data point is nearly a linear combination of the existing ones, so it hardly changes the singular vectors in \eq{eq:J-svd}, but only rescales $\lambda_k^2$ by a factor of order $(1 + N_w^{-1})$, with $N_w$ the number of existing data points.
After saturation, $\kappa$ therefore grows logarithmically with $N_w$.
In practice, we fit $\kappa = a + b \ln N_w$ and find the smallest $N_w$ beyond which all values of $\kappa$ agree with the fit within $0.5\%$; the saturation point is taken as the next $N_w$.
The resulting data points for each type of observable are summarized in Table~\ref{tab:pseudodata-summary}.

\begin{table*}[b]
\centering
\caption{Summary of the pseudodata in each set. Each real or imaginary component at one kinematic point is counted as one scalar datum.}
\label{tab:pseudodata-summary}
\vspace{2mm}
\begin{tabular}{p{0.15\textwidth}p{0.26\textwidth}c@{\hspace{1.2em}}p{0.46\textwidth}}
\hline\hline
Pseudodata & Included quantities & Points~ & Kinematic coverage\\
\hline
DVCS integrals
& $\Re\H(\xi, \mu_0)$, $\Im\H(\xi, \mu_0)$
& \phantom{0}66
& 33 values of $\xi \in [0.1,0.9]$; $\mu_0 = 2~\GeV$
\\
Evolved DVCS
& $\Re\H(\xi, Q)$, $\Im\H(\xi, Q)$  & 330
& same $\xi$ points,
5 values of $Q \in [2,\, 10]~\GeV$
\\
Mesoproduction
& $\Re I^{\gamma\gamma}(\xi, \cos\theta)$, $\Im I^{\gamma\gamma}(\xi, \cos\theta)$
& 528
& same $\xi$ points,
8 values of $\cos\theta \in [0.182, 0.9]$; $\mu_0 = 2~\GeV$
\\
Photoproduction
& $\Re I^{\gamma\pi}(\xi, \cos\theta)$, $\Im I^{\gamma\pi}(\xi, \cos\theta)$
& 792
& same $\xi$ points,
12 values of $\cos\theta \in [-0.9, 0.9]$; $\mu_0 = 2~\GeV$
\\
\hline\hline
\end{tabular}
\end{table*}

\begin{figure}[t]
     \centering     
     \includegraphics[scale=1.0]{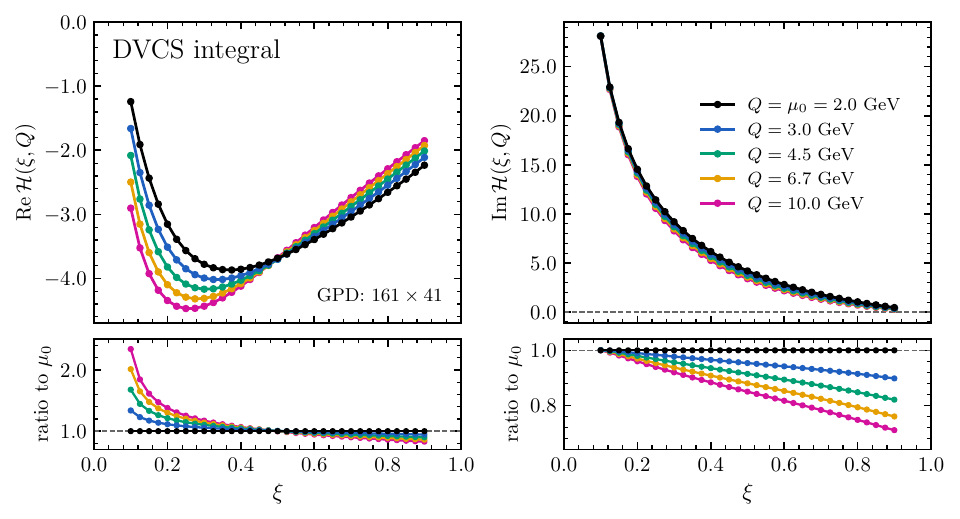}
     \caption{Noise-free pseudodata for the real (left) and imaginary (right) parts of the DVCS integral $\H(\xi, Q)$,
     evaluated for the GK GPD implemented on the grid $N_x \times N_\xi = 161 \times 41$.
     The markers show the 33 points of $\xi$ at each of the five $Q$ values.
     Ratios of the evolved integrals to $Q = \mu_0$ are shown in the bottom panels.}
\label{fig:dvcs-integral}
\end{figure}

For the DVCS integrals $\H(\xi, \mu_0)$ in \eq{eq:obs-mat2-dvcs} at the fixed scale $\mu_0 = 2~\GeV$, we place the data points of $\xi$ exactly on the GPD grid nodes, so that each $\H(\xi_j, \mu_0)$ depends only on the pixels of the $j$-th $\xi$ slice and different $\xi$ values provide independent constraints.
We therefore use all grid values of $\xi_j \in [0.1, 0.9]$, outside which the GK GPD varies too rapidly around the ridge $|x| = \xi$ and a finer resolution of the linear grid is needed to reduce numerical integration errors.
This gives 33 $\xi$ values for both $\Re\H$ and $\Im\H$, 
whose central values $\bar w_a$ are shown by the black curves in \fig{fig:dvcs-integral}.

\begin{figure}[t]
     \centering
     \includegraphics[scale=1.0]{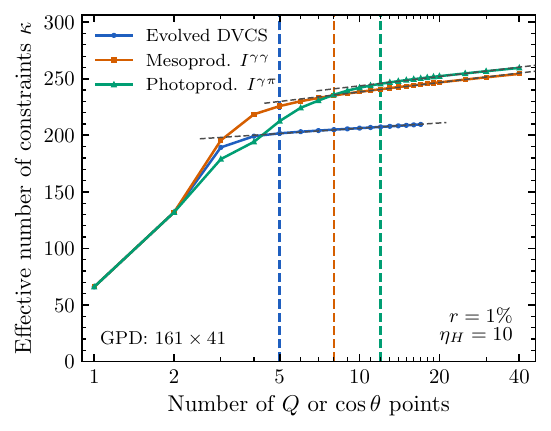}
     \caption{Effective numbers of constraints, $\kappa$,
     as functions of the number of $Q$ values for the evolved DVCS integrals,
     or of $\cos\theta$ values for the mesoproduction and photoproduction NSIs,
     with $r = 0.01$ and $\eta_H = 10$.
     The horizontal axis is in logarithmic scale. Vertical dashed lines indicate the saturation points (at 5, 8, and 12), 
     beyond which $\kappa$ grows logarithmically,
     and the thin dashed gray lines show the corresponding fits.}
\label{fig:fisher-saturation}
\end{figure}

Including the evolution as in \eq{eq:evo-dvcs-mat}, we keep the same $\xi$ points and sample $Q$ logarithmically in the range $[2, 10]~\GeV$, where the lower end coincides with $\mu_0$.
As shown in \fig{fig:fisher-saturation}, $\kappa$ saturates at around five $Q$ values, with $\kappa = 202$, which yields $5 \times 33 \times 2 = 330$ data points for $\Re\H$ and $\Im\H$.
Their noise-free pseudodata are shown in \fig{fig:dvcs-integral} together with the unevolved ones.
Over this $Q$ range, evolution changes the DVCS integrals by up to about $10\%$ at most $\xi$ points.

\begin{figure}[t]
     \centering     
     \includegraphics[scale=0.95]{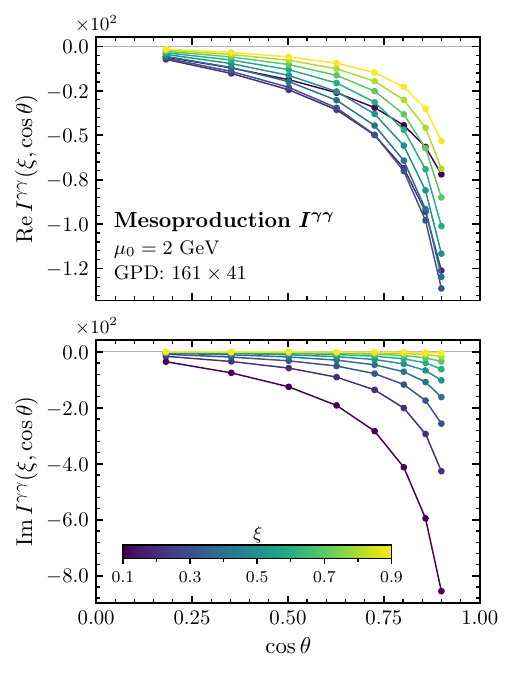}
     \includegraphics[scale=0.95]{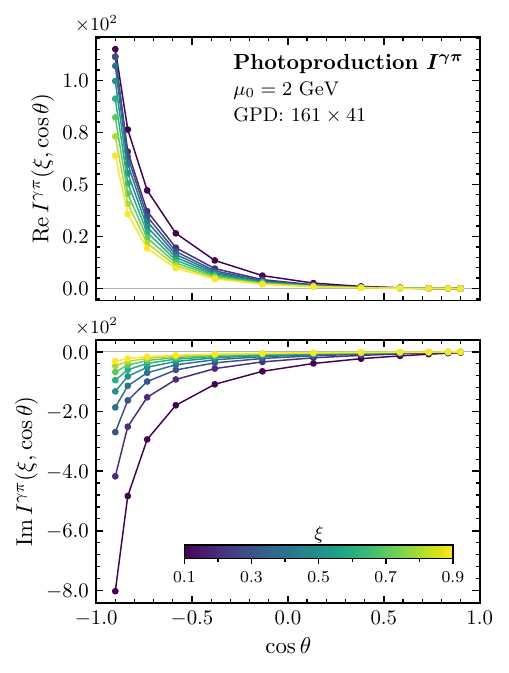}
     \caption{Noise-free pseudodata for the mesoproduction (left) and photoproduction (right) NSIs, evaluated for the GK GPD implemented on the grid $N_x \times N_\xi = 161 \times 41$. Upper and lower panels are the real and the imaginary parts, respectively. The data are shown for only a subset of the 33 $\xi$ points for visibility.}
\label{fig:nsi-data}
\end{figure}

For the two NSIs in \eq{eq:nsi-prod}, we keep the same $\xi$ points and choose $\cos\theta \in [-0.9, 0.9]$, sampled according to the distribution $1/(1-\cos^2\theta)$. 
This places more points at large $|\cos\theta|$, where the NSIs vary most rapidly.
Since $H^{u+}$ is odd in $x$ and the asymptotic DA is symmetric under the interchange $z \leftrightarrow 1 - z$, the integral $I^{\gamma\gamma}$ in \eq{eq:nsi-meso-x} is odd in $\cos\theta$.
Consequently, only $\cos\theta > 0$ carries independent information for mesoproduction, and we sample $I^{\gamma\gamma}$ in the range $(0, 0.9]$ only.
The effective constraints $\kappa$ are shown in \fig{fig:fisher-saturation} as functions of the number of $\cos\theta$ points.
Compared with the evolved DVCS integrals, the $\kappa$'s of both NSIs grow to larger values and saturate later, at around 8 and 12 $\cos\theta$ points for $I^{\gamma\gamma}$ and $I^{\gamma\pi}$, with $\kappa = 235$ and $246$, respectively, which yields $8 \times 33 \times 2 = 528$ and $12 \times 33 \times 2 = 792$ data points for their real and imaginary parts.
Their noise-free pseudodata are shown in \fig{fig:nsi-data}, where the nontrivial $\cos\theta$ dependence at each fixed $\xi$ is clearly visible.

\clearpage

\section{Results}
\label{sec:results}

Having described the generation of the pseudodata in \sec{sec:data}, we now perform the inversion using the exact Gaussian posterior in \sec{sec:bayesian}, first at the GPD level (\sec{sec:results-gpd}) and then at the DD level (\sec{sec:results-dd}).
Throughout this section, we mainly use $r = 0.01$ but examine the variation of $r$ in \sec{sec:results-robust}.
Each inversion of the GPD is defined by 
(a) whether it is performed at the GPD or the DD level, 
(b) the grid choice in the case of the latter, and 
(c) the combination of observables from Table~\ref{tab:pseudodata-summary}, being one of the following categories:
\begin{enumerate}
\item[(1)] the DVCS integrals at the fixed scale $\mu_0 = 2~\GeV$, to be referred to as ``DVCS'', with $N_{\rm pts} = 66$ data points,
\item[(2)] the evolved DVCS integrals, including those at $\mu_0$, to be referred to as ``DVCS$(Q)$'', with $N_{\rm pts} = 330$, and
\item[(3)] the evolved DVCS integrals and both NSIs, to be referred to as ``DVCS$(Q)$ + NSIs'', with $N_{\rm pts} = 330 + 528 + 792 = 1650$.
\end{enumerate}
In each inversion, we denote the posterior mean of the GPD pixels by $\langle H^{u+}_{ij} \rangle$, which is given by $\bm h$ in \eq{eq:posterior-mean-g} at the GPD level, and by $\bm R \bm d$ with \eq{eq:posterior-mean-d} at the DD level.
In both cases, we take zero prior means, $\bm h_0 = \bm 0$ and $\bm d_0 = \bm 0$, to avoid the bias terms in \eqs{eq:posterior-mean-g}{eq:posterior-mean-d}.

To quantify how well the ground truth is reconstructed in each inversion, we define the relative deviation of the posterior mean from the truth,
\beq[eq:rel-error]
    (\delta_H)^2 =
    \frac{\sum_{i,j} \pp{ \langle H^{u+}_{ij} \rangle - \widebar H^{u+}_{ij} }^2}
         {\sum_{i,j} \pp{ \widebar H^{u+}_{ij} }^2},
\eeq
where the sum runs over the pixels with $x_i > 0$ and $\xi_j \in [0.1, 0.9]$, corresponding to the $\xi$ range covered by the pseudodata.
To check how well the posterior mean describes the pseudodata, we also evaluate its $\chi^2$ according to
\beq[eq:posterior-mean-chi2]
	\chi^2 
	= \bigbb{ \bm w - \bm T (\vv{ \bm\epsilon }) }^\T \, 
		\bm \Sigma^{-1} \,
		\bigbb{ \bm w - \bm T (\vv{ \bm\epsilon }) }
	= \sum_a \frac{\bigpp{ w_a - A_{am} \vv{ \epsilon_m } - B_a }^2}{
		r^2 \, \widebar w_a^2},
\eeq
where $\langle \bm\epsilon \rangle$ is the vector of $\langle H^{u+}_{ij} \rangle$, 
and the sum runs over the $N_{\rm pts}$ data points included in the inversion.
We will examine inversions both with statistical fluctuations and without.
In the former case, random noise is injected into the data,
$w_a = \widebar w_a + r \widebar w_a \, \delta w_a$ with $\delta w_a$ drawn from $\N(0, 1)$,
before the inversion and used also in computing the $\chi^2$.
In the latter case, $w_a = \widebar w_a$ without including any statistical fluctuation in the data,
we should expect $\chi^2 / N_{\rm pts} \ll 1$.

The results are summarized in Table~\ref{tab:results-chi2} for the noise-free pseudodata, 
and in Table~\ref{tab:results-chi2-noisy} for the noisy pseudodata with 100 replicas generated in each data category.
In the following, we first look into the GPD-level inversion and then the DD level.

\begin{table*}[h]
\centering
\caption{Relative deviation $\delta_H$ (in \%) of the posterior mean from the ground truth, with the $\chi^2$ in \eq{eq:posterior-mean-chi2} per data point, $\chi^2 / N_{\rm pts}$, given in parentheses, for each inversion of the noise-free pseudodata with $r = 0.01$. The GPD grid is fixed to $N_x \times N_\xi = 161 \times 41$. For each inversion at the DD level, the grid $N_u \times N_v$ is also given in parentheses, with the prior width $\eta_D = 63.4$, $76.0$, and 90.8 of the DD pixels for $N_u = 81$, 115, and 163, respectively.}
\label{tab:results-chi2}
\vspace{2mm}
\setlength{\tabcolsep}{12pt}
\begin{tabular}{lccccc}
\hline\hline
Pseudodata & $N_{\rm pts}$ & Inversion at GPD & DD ($81 \times 81$) & DD ($115 \times 115$) & DD ($163 \times 163$) 
\\ \hline
DVCS & \phantom{00}66 & 97.9 ($1.9\times10^{-6}$) & 61.6 ($8.0\times10^{-8}$) & 67.1 ($9.5\times10^{-8}$) & 72.3 ($1.1\times10^{-7}$) 
\\
DVCS$(Q)$ & \phantom{0}330 & 88.1 ($5.5\times10^{-3}$) & 15.7 ($2.5\times10^{-4}$) & 19.5 ($3.6\times10^{-4}$) & 22.5 ($4.4\times10^{-4}$)
\\
DVCS$(Q)$ + NSIs & 1650 & \phantom{0}1.9 ($1.3\times10^{-5}$) & \phantom{0}5.3 ($4.3\times10^{-5}$) & \phantom{0}5.9 ($6.4\times10^{-5}$) & \phantom{0}3.9 ($1.4\times10^{-5}$) 
\\ \hline\hline
\end{tabular}
\end{table*}

\begin{table*}[h]
\centering
\caption{As in Table~\ref{tab:results-chi2}, but for noisy pseudodata with $r = 0.01$: the mean and standard deviation of $\delta_H$ (in \%) over $100$ replicas, with the mean $\chi^2 / N_{\rm pts}$ given in parentheses. The prior widths are the same as in Table~\ref{tab:results-chi2}.}
\label{tab:results-chi2-noisy}
\vspace{2mm}
\setlength{\tabcolsep}{12pt}
\begin{tabular}{lccccc}
\hline\hline
Pseudodata & $N_{\rm pts}$ & Inversion at GPD & DD ($81 \times 81$) & DD ($115 \times 115$) & DD ($163 \times 163$) 
\\ \hline
DVCS & \phantom{00}66 & 97.9 $\pm$ 0.0 (0.00) & 61.6 $\pm$ 0.1 (0.00) & 67.1 $\pm$ 0.1 (0.00) & 72.3 $\pm$ 0.0 (0.00) 
\\
DVCS$(Q)$ & \phantom{0}330 & 93.6 $\pm$ 1.4 (0.36) & 57.0 $\pm$ 7.0 (0.45) & 55.2 $\pm$ 6.2 (0.42) & 50.7 $\pm$ 5.4 (0.40) 
\\
DVCS$(Q)$ + NSIs & 1650 & 51.8 $\pm$ 3.2 (0.70) & 67.3 $\pm$ 6.6 (0.79) & 65.1 $\pm$ 5.4 (0.78) & 54.6 $\pm$ 5.1 (0.75) 
\\ \hline\hline
\end{tabular}
\end{table*}

\subsection{Inversion at GPD level}
\label{sec:results-gpd}

For the inversion at the GPD level, we take the GPD pixels $H^{u+}_{ij}$ with $x_i > 0$ as the free parameters,
which amount to $n_G = (N_x - 1)/2 \times N_\xi = 80 \times 41 = 3280$ in total,
and use the Gaussian prior $\mathcal{N}(0, \eta_H^2)$ with $\eta_H = 10$ for each pixel independently,
\beq[eq:V0-gpd]
	\bm V_0 = \eta_H^2\, \mathbf{1}.
\eeq
This setup does not guarantee polynomiality, and the pixels at different $\xi$ are independent.
The ground truth is simply given by the GK GPD at the grid nodes, which also generate the pseudodata as in \sec{sec:data}.
Since every observable involves the GPD at a single $\xi$ value, the matrix $\bm J$ in \eq{eq:J-def} is block diagonal in $\xi$, and the SVD in \sec{ssec:svd-g} can be performed for each $\xi$ slice separately, with $n_G / N_\xi = 80$ directions per slice.

\subsubsection{Quality of the inversion}
\label{sec:results-gpd-fit}

We next would like to express $\delta_H$ and $\chi^2$ in terms of the SVD analysis results in \sec{ssec:svd-g}.
In our setup with \eq{eq:V0-gpd} for the GPD pixel prior, we have the scaling $\bm\epsilon = \sqrt{\bm V_0} \bm z = \eta_H \bm z$;
additionally with \eq{eq:data-sigma} for the data correlation, the singular values scale as $\lambda_k\ \propto\ \eta_H / r$.
With $\bm h_0 = \bm 0$, \eq{eq:post-mean-z-svd-2} gives the deviation of the posterior mean from the truth,
\beq[eq:dev-svd]
	\bm\mu_z - \bar{\bm z}\,
	=\, - \sum_{k = 1}^{n_G} \frac{ \bar z_k }{ 1 + \lambda_k^2 } \, \bm v^{(k)}\,
	+\, \sum_{k = 1}^{r_A} \frac{ \lambda_k \, \delta w_k }{ 1 + \lambda_k^2 } \, \bm v^{(k)},
\eeq
where $\bar z_k = \bm v^{(k)\T} \bar{\bm z}$ is the component of the truth GPD, and $\delta w_k = \bm u^{(k)\T} \delta\bm w$ is the component of the data noise in the $k$-th singular direction.
Since $\delta\bm w \sim \N(0, \mathbf{1})$ and the $\bm u^{(k)}$ are orthonormal, $\delta w_k$ are independent standard normal variables.
Averaging over the noise and using the orthonormality of $\bm v^{(k)}$, we obtain
\beq[eq:deltaH-svd-noise]
    \bigvv{ (\delta_H)^2 }\, 
    =\, \frac{1}{\sum_{k = 1}^{n_G} \bar z_k^2} \biggbb{
    	\sum_{k = 1}^{n_G} \frac{\bar z_k^2 }{ (1 + \lambda_k^2)^2 }\,
    +\, \sum_{k = 1}^{r_A} \frac{\lambda_k^2 }{ (1 + \lambda_k^2)^2 }
	},
\eeq
which applies to each single $\xi$ point in the data, and where we have used $\vv{\delta w_k} = 0$ and $\vv{\delta w_k \delta w_{k'}} = \delta_{kk'}$.
Similarly, the $\chi^2$ in \eq{eq:posterior-mean-chi2} is the squared norm of the residuals normalized by the data uncertainties,
\begin{align}
	\bm\Sigma^{-1/2} (\bm w - \bm B - \bm A \bm h) 
	&= \bm J (\bar{\bm z} - \bm\mu_z) + \delta\bm w 
	= \sum_{k = 1}^{N_{\rm pts}} \bm u^{(k)} \,
		\frac{\lambda_k \bar z_k + \delta w_k}{1 + \lambda_k^2} ,
\end{align}
where we have used \eq{eq:J-svd} with the convention that $\lambda_k = 0$ for $k > r_A$ and completed the singular vectors $\{\bm u^{(k)}\}_{k = 1}^{r_A}$ to an orthonormal basis $\{\bm u^{(k)}\}_{k = 1}^{N_{\rm pts}}$ of the observable space.
After averaging over the noise, the $\chi^2$ then becomes
\beq[eq:chi2-svd]
    \bigvv{ \chi^2 }\,
	=\, \sum_{k = 1}^{r_A} \frac{ \lambda_k^2 \, \bar z_k^2 }{ (1 + \lambda_k^2)^2 }\
	+\ \sum_{k = 1}^{N_{\rm pts}} \frac{1}{ (1 + \lambda_k^2)^2 }.
\eeq
The second terms of both \eqs{eq:deltaH-svd-noise}{eq:chi2-svd} come from the noise.

Comparing \eqs{eq:deltaH-svd-noise}{eq:chi2-svd}, we see opposite roles of the numerator $\lambda_k^2$ factors in the GPD recovery and in the $\chi^2$.
Without inclusion of noise, $(\delta_H)^2$ is dominated by the directions that are not well constrained, particularly the shadow GPDs, with little contribution from the well constrained directions.
In contrast, the $\chi^2$ is mostly carried by the moderately constrained directions, with $\lambda_k^2 \sim 1$.
For the noise terms, on the other hand, the roles are reversed.
\footnote{Note that the sums in \eq{eq:deltaH-svd-noise} run over the $n_G$ directions $\bm v^{(k)}$ in the GPD space, and those in \eq{eq:chi2-svd} over the $N_{\rm pts}$ directions $\bm u^{(k)}$ in the observable space. They share only $k \leq r_A$, with the same $\lambda_k$. The directions with $k > r_A$ differ: in \eq{eq:deltaH-svd-noise} they span the null space of $\bm A$, which contains most of the shadow GPDs, whereas in \eq{eq:chi2-svd} they are the combinations of data that no GPD can change.}
Along a data direction $\bm u^{(k)}$ with $\lambda_k \gg 1$, the pixels fit the noise and remove it from the $\chi^2$, while along one with $\lambda_k \ll 1$ (or $k > r_A$), the noise remains.
Hence $\chi^2 / N_{\rm pts}$ of noisy data is below $1$ by about the fraction of the data directions with large $\lambda_k$ (which notably is not the same as the number of constrained pixel directions).
In either case, with or without noise, an accurate fit to the data does not imply an accurate GPD.

\begin{figure}[t]
    \centering
    \includegraphics[width=0.88\textwidth]{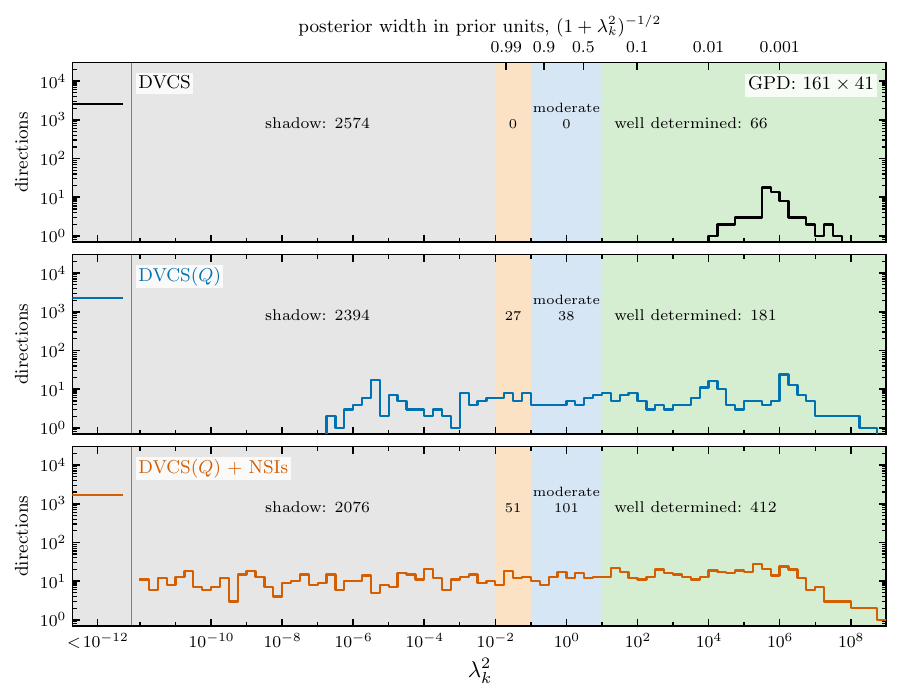}
    \vspace*{-0.3cm}
    \caption{Distribution of the squared singular values $\lambda_k^2$ of $\bm J$ in \eq{eq:J-svd} at the GPD level, with $r = 0.01$ and $\eta_H = 10$, for the three combinations of pseudodata, counting the 80 singular directions of each of the 33 $\xi$ slices with data (2640 directions in total). The histograms are binned logarithmically in $\lambda_k^2$, and the directions with $\lambda_k^2 < 10^{-12}$, including those with $\lambda_k = 0$, are collected in the leftmost bin. The shaded regions indicate (from left to right): the shadow GPD space ($\lambda_k^2 \leq 0.01$, gray), the directions with $0.01 < \lambda_k^2 \leq 0.1$ (orange), the moderately constrained directions ($0.1 < \lambda_k^2 < 10$, blue), and the well constrained directions ($\lambda_k^2 \geq 10$, green), with the number of directions in each region indicated. The top axis gives the posterior width along $\bm v^{(k)}$ in units of the prior width, $(1 + \lambda_k^2)^{-1/2}$.}
\label{fig:gpd-level-svd-spectrum}
\end{figure}

\begin{figure*}[t]
    \centering
    \includegraphics[scale=0.8]{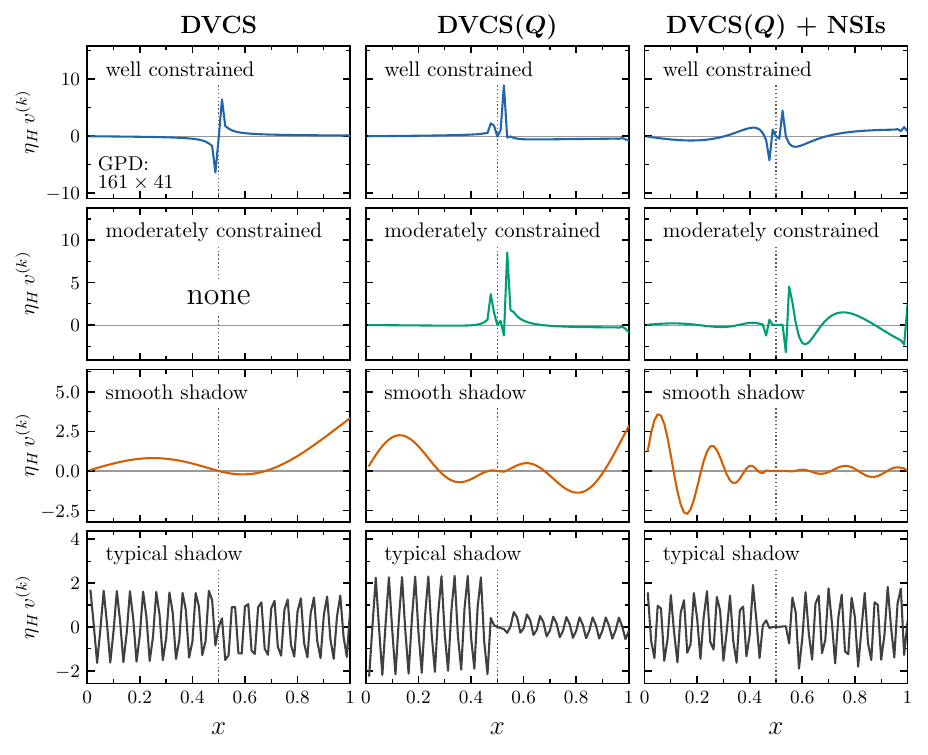}
    \vspace*{-0.5cm}
    \caption{Representative singular directions $\bm v^{(k)}$ of the inversion at the GPD level at $\xi = 0.5$ for the three combinations of pseudodata (columns), with $r = 0.01$ and $\eta_H = 10$, each rescaled by the prior width $\eta_H$.
    From top to bottom, the rows show a well constrained direction with $\lambda_k^2 > 100$, a moderately constrained one with $\lambda_k^2$ closest to $1$ in $[0.1, 10]$, the smoothest shadow GPD $\bm s$, and a more typical, strongly oscillating, shadow GPD, the last two with $\lambda_k^2 \leq 0.01$.
    For the ``DVCS'' inversion, there are no moderately constrained directions.
    The dotted lines mark $x = \xi$.}
\label{fig:gpd-level-directions}
\end{figure*}

The distribution of the singular values is shown as a histogram of $\lambda_k^2$ in \fig{fig:gpd-level-svd-spectrum} for each category of the observables.
These are divided into the characteristic regions defined below \eq{eq:post-mean-z-svd-2}, and examples of the singular vectors from these regions are illustrated in \fig{fig:gpd-level-directions}.
For the DVCS integrals at a single scale, every direction is either well constrained or in the shadow GPD space, with exactly one well constrained direction per data point, sharply concentrated around the ridge, so that each data point provides an independent and important constraint on the pixels.
The evolution and the NSIs instead produce singular values spread over many orders of magnitude, with a fraction of their directions only moderately constrained.
The NSIs add by far the most well constrained directions, which are less sharply peaked at $x = \xi$ and extend over a wider range of $x$.
Even with these, however, most directions (around 80\% in \fig{fig:gpd-level-svd-spectrum}) remain in the shadow GPD space.

These findings are in agreement with the results in Tables~\ref{tab:results-chi2} and \ref{tab:results-chi2-noisy}.
Without noise, $\chi^2/N_{\rm pts}$ never exceeds $10^{-2}$, so all the posterior means describe the pseudodata perfectly, while $\delta_H$ varies from  $\approx 2\%$ to nearly 100\%.
With noise, $\chi^2/N_{\rm pts}$ stays below 1, in agreement with \eq{eq:chi2-svd}.
For the DVCS integrals, all the data directions are well constrained in \fig{fig:gpd-level-svd-spectrum},
so that every data point is fitted by the free pixels, giving $\chi^2 / N_{\rm pts} \approx 0$ with or without noise.
With the NSIs, many more pixel directions are constrained,
but most of the data directions have $\lambda_k \leq 0.1$,
along which no GPD within the prior width can shift the data appreciably.
This is because while the $\cos\theta$ distributions of the NSIs provide nontrivial constraints on the $x$ dimension of the GPD pixels, 
at neighboring $\cos\theta$, the NSIs depend on the GPD in nearly the same way.
Their noise therefore remains in the $\chi^2$, giving $\chi^2 / N_{\rm pts} \approx 0.7$.

\begin{figure*}[htbp]
    \centering
    \includegraphics[scale=1.0]{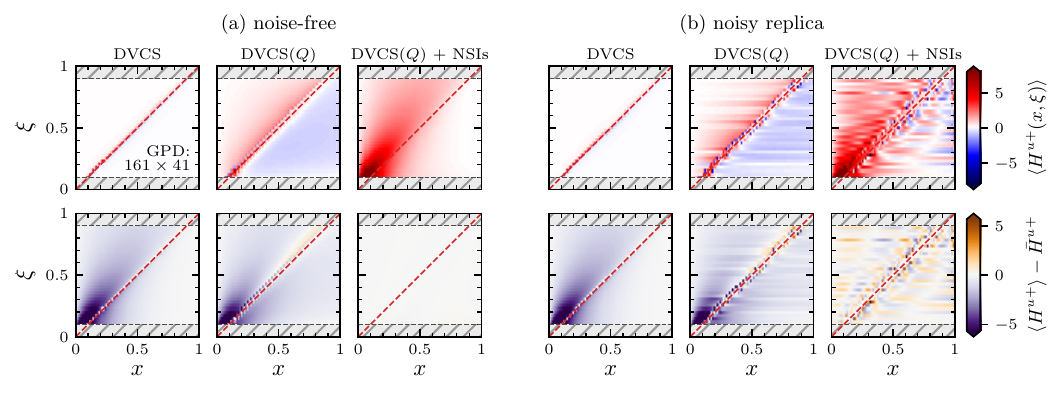}
    \caption{Posterior of the inversion at the GPD level for the three combinations of pseudodata, with $r = 0.01$ and $\eta_H = 10$, for (a) the noise-free pseudodata and (b) one replica of noisy pseudodata. The upper row shows the posterior mean $\langle H^{u+}(x, \xi) \rangle \equiv \langle H^{u+}_{ij} \rangle$, and the lower row shows its difference from the ground truth, $\langle H^{u+} \rangle - \bar{H}^{u+}$, as density plots, with the same color scale in (a) and (b). The data cover only $\xi \in [0.1, 0.9]$ (between the gray dashed lines); outside this range (hatched), the posterior mean stays at the prior mean (which is zero), for every data set, so its difference from the truth is minus the truth. Only the $x > 0$ parts are shown, and the red dashed lines mark $x = \xi$.}
\label{fig:gpd-level-maps}
\end{figure*}

\begin{figure*}[t]
    \centering
    \includegraphics[scale=1.0]{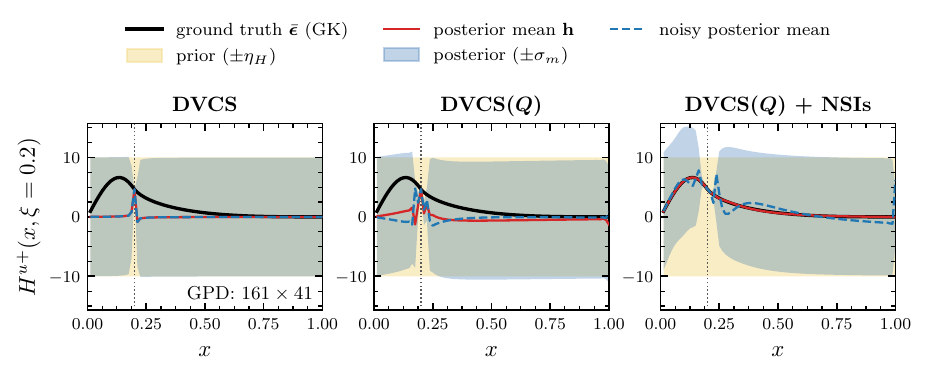}
    \vspace*{-0.5cm}
    \caption{Posterior of the inversion at the GPD level at $\xi = 0.2$
    for the three combinations of pseudodata (columns), with $r = 0.01$ and $\eta_H = 10$.
    The black solid curves show the ground truth $\bar{\bm\epsilon}$ (GK model), 
    the red solid curves the posterior mean $\bm h$ from the noise-free pseudodata, and the blue dashed curves the posterior mean from the noisy pseudodata in \fig{fig:gpd-level-maps}(b).
    The bands show the pointwise one-standard-deviation ranges of the GPD in the prior and the posterior.
    The vertical dotted lines mark $x = \xi$. Only $x > 0$ is shown.}
\label{fig:gpd-level-bands-shadow}
\end{figure*}

The posterior means and their differences from the ground truth are shown in \fig{fig:gpd-level-maps} for all three observable combinations, and are also shown in $x$ space at $\xi = 0.2$ in \fig{fig:gpd-level-bands-shadow}.
With the DVCS data alone, the posterior mean only departs from the prior mean (which is at zero) along the ridge $x = \xi$, where $\Im \H$ directly samples the GPD and the two well constrained directions of each slice are localized.
Elsewhere the difference from the truth is as large as the GPD itself, giving $\delta_H = 98\%$ (cf.~Table~\ref{tab:results-chi2}).
Including the evolution adds well constrained directions,
but the posterior mean is only moderately improved, still with $\delta_H = 88\%$, and the GPD away from the ridge remains mostly undetermined.
In contrast, the NSIs bring the posterior mean close to the truth over the whole region covered by the data,
reducing $\delta_H$ to $1.9\%$.

It may be surprising that the ``DVCS$(Q)$ + NSIs'' data recover the GPD so accurately, even though a large portion of the singular directions lie in the shadow subspace.
To understand this, it is important to notice from \eq{eq:dev-svd} or \eq{eq:deltaH-svd-noise} that the difference of the posterior mean from the ground truth depends not only on the data, which determine the singular values and vectors and thereby the observed and shadow GPD subspaces, {\it but also on the ground truth $\bar{\bm z}$}, mainly through its overlap with the shadow GPD subspace.
Along the shadow GPDs, the posterior mean stays at the prior mean by the last term of \eq{eq:post-mean-z-svd-2}.
Without the NSIs, the large shadow GPD subspace contains most of the GK model, whereas with the NSIs, the truth lies almost entirely within the observed GPD subspace.
We will see this more transparently in \sec{sec:results-gpd-proj}.

The posterior uncertainty band, on the other hand, stays close to the prior width in all cases, except for a narrow dip around $x = \xi$, even when the posterior mean agrees well with the truth.
Defined by \eq{eq:sigma-m} point by point in $(x, \xi)$, it is dominated by the shadow GPD subspace, as argued below \eq{eq:cov-z-g-svd-2}, and mostly spanned by the shadow GPDs that strongly oscillate with $x$ as illustrated in the last row of \fig{fig:gpd-level-directions}.
These shadow GPDs depend sensitively on the choice of the $x$ grid and can be largely removed by lowering the resolution.
However, there are also smooth shadow GPDs, which are much less sensitive to the grid resolution.
For example, the smoothest one of these, to be denoted by $\bm{s}$, is shown in the third row of \fig{fig:gpd-level-directions}, obtained by minimizing the second-order difference in $x$ within the shadow GPD subspace of each inversion.
It lies away from $x = \xi$, but otherwise depends strongly on the observables.
For the DVCS integrals, with or without the evolution, it varies slowly over most of the $x$ range, so that broad deformations of the GPD at fixed $\xi$ are left unconstrained.
Once the NSIs are included, even this smoothest shadow GPD oscillates rapidly in $x$.
Hence, the remaining ambiguity is of high frequency in $x$ only with the NSIs, which also explains why a smooth GPD like the GK model is well reconstructed in this case.

\begin{figure*}[t]
    \centering
    \includegraphics[scale=1.0]{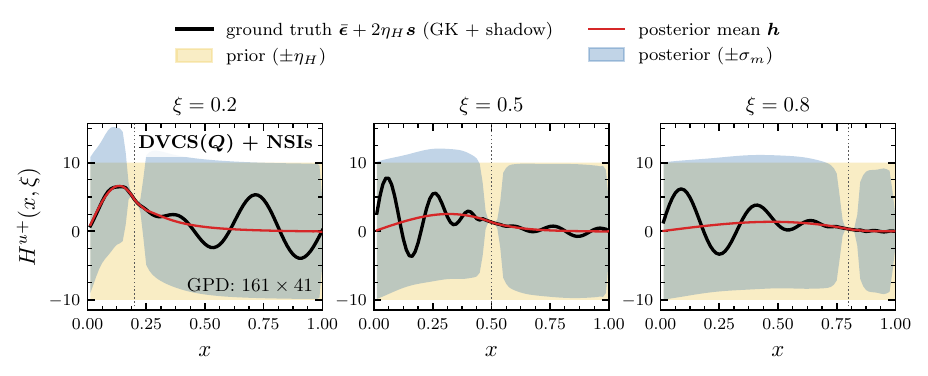}
    \vspace*{-0.5cm}
    \caption{Inversion at the GPD level of the ``DVCS$(Q)$ + NSIs'' pseudodata generated from a ``shadow-modified'' ground truth (see text), at $\xi = 0.2$, $0.5$, and $0.8$ (columns), with $r = 0.01$ and $\eta_H = 10$. The bands and the posterior mean are as in \fig{fig:gpd-level-bands-shadow},
    and the black curves show the new truth.}
\label{fig:shadow-truth}
\end{figure*}

That also means that if we started with a ground-truth model that differs from the GK model by one of the shadow GPDs, it would not be recovered correctly.
To test this, we add twice the smooth shadow $\bm s$ of the ``DVCS$(Q)$ + NSIs'' data in each $\xi$ slice to the GK model to get a modified ground truth,
$\bar{\bm\epsilon} + 2 \eta_H \bm s$, regenerate the pseudodata, and repeat the inversion.
Since $\bm A \bm s \approx \bm 0$ by definition, neither the pseudodata nor the posterior changes appreciably. 
The new posterior mean still follows the unmodified GK model and completely misses the added shadow, as shown in \fig{fig:shadow-truth},
where the shadow is concentrated at $x > \xi$ for small $\xi$ and at $x < \xi$ for large $\xi$. 
The data thus cannot distinguish the two truths. The accurate recovery of the GK model with the NSIs therefore relies in part on its smoothness, 
that is, on its small overlap with the shadow GPD subspace of the data we use.

Now let us examine how the data fluctuations around their central values affect the inversion.
Since the posterior covariance in \eq{eq:posterior-cov-g} does not depend on $\bm w$,
only the posterior mean responds to the fluctuations.
Figure~\ref{fig:gpd-level-maps}(b) shows the inversion of one replica of the noisy pseudodata.
For the ``DVCS'' inversion, the posterior mean is almost unchanged,
because all its directions have either $\lambda_k^2 \geq 10$ or $\lambda_k^2 \leq 0.01$,
for which the noise term in \eq{eq:deltaH-svd-noise} is negligible.
For the ``DVCS$(Q)$'', $\delta_H$ increases only moderately from $88\%$ to $94\%$.
More dramatically, once the noise is added to the NSIs, the previously well recovered posterior mean develops oscillations in $x$,
most prominently around $x = \xi$.
Averaged over the data noise, $\delta_H$ becomes $(52 \pm 3)\%$ for the ``DVCS$(Q)$ + NSIs'' data (Table~\ref{tab:results-chi2-noisy}), compared with $1.9\%$ for the noise-free pseudodata (Table~\ref{tab:results-chi2}).
In \fig{fig:gpd-level-bands-shadow}, we show the posterior mean for the noisy inversion in \fig{fig:gpd-level-maps}(b) by the dashed curves.
As expected from \eq{eq:noise-ratio}, the fluctuations follow the moderately constrained directions illustrated in \fig{fig:gpd-level-directions}
and stay within the posterior bands.

It is important to remember that the noise term in \eq{eq:deltaH-svd-noise}
contributes mainly through the moderately constrained directions, 
concentrating in the window $0.1 \lesssim \lambda_k^2 \lesssim 10$ of \fig{fig:gpd-level-svd-spectrum}.
This means that the impact of noise on $\delta_H$ is not reduced by increasing the precision of the data.
Since $\lambda_k \propto \eta_H / r$, more precise data only shift the spectrum to larger $\lambda_k$,
which hardly changes the sum $\sum_k \lambda_k^2 / (1 + \lambda_k^2)^2$
for the evolved DVCS integrals with or without the NSIs, whose spectra are almost flat in $\log\lambda_k^2$;
for the ``DVCS'' inversion, all singular values lie above this window and the noise term is negligible.
Also, since our choice of data points has already saturated the $\kappa$ of each observable type,
adding more data points of $Q$ to the ``DVCS$(Q)$'' or $\cos\theta$ to both NSIs 
does not help to reduce the noise impact, either.
In contrast, the prefactor $1 / \sum_k \bar z_k^2 = \eta_H^2 / |\bar{\bm\epsilon}|^2$ in \eq{eq:deltaH-svd-noise}
makes the noise contribution to $\delta_H$ grow nearly in proportion to $\eta_H$.
Hence the noise in the posterior mean is controlled by the prior width rather than by the precision or the amount of data,
which we have verified numerically.

The data could possibly be de-noised by choosing suitable off-diagonal elements of the covariance matrix 
in the prior pixel distribution,
which would then bring the inversion closer to the noise-free scenario
and allow the high-dimensional observed GPD subspace of the NSIs
to pin down a large portion of the GPD through the posterior mean.
We leave that option to future study.

\subsubsection{Constrained and shadow components}
\label{sec:results-gpd-proj}

As seen from \eqs{eq:post-mean-z-svd-2}{eq:cov-z-g-svd-2}, the pointwise uncertainty is determined by the poorly constrained directions,
whereas the posterior mean follows the truth only along the well constrained ones.
These two thus cannot be inferred from each other, as we have seen in \fig{fig:gpd-level-bands-shadow},
except for the narrow region near the ridge.
This fact can be seen more transparently from the projectors $\widehat P_\ob$ and $\widehat P_\sh$ in \eqs{eq:P-ob}{eq:P-sh}.

\begin{figure*}[t]
    \centering
    \includegraphics[scale=0.95]{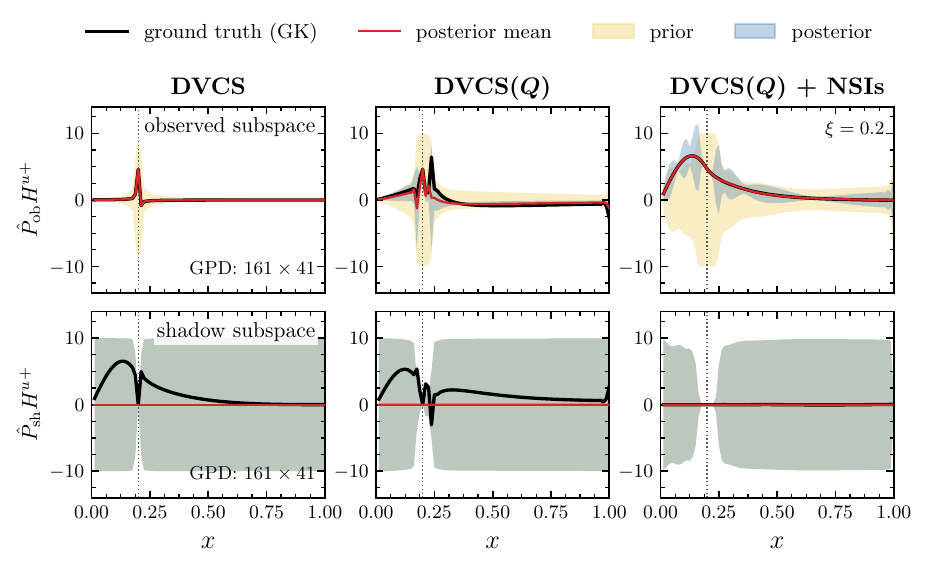}
    \vspace*{-0.5cm}
    \caption{Decomposition of the full posterior at the GPD level in \fig{fig:gpd-level-bands-shadow}
    into the observed subspace (upper row) and the shadow GPD space (lower row)
    for the three combinations of pseudodata, with $r = 0.01$ and $\eta_H = 10$.
    The prior and posterior bands coincide in the lower row.
    The vertical dotted lines mark $x = \xi$.}
\label{fig:gpd-level-subspaces}
\end{figure*}

Though defined with $\bm v^{(k)}$ in the $\bm z$ space, 
these projectors also act directly on the GPD pixels $\bm h$,
since $\bm h = \sqrt{\bm V_0} \bm z = \eta_H \bm z$.
Since they are built from the eigenvectors of $\bm V_z$, they commute with $\bm V_{\rm post}$,
and the posterior splits into the observed and shadow subspaces without cross terms,
\beq[eq:post-split]
	\bm h = \widehat P_\ob \bm h + \widehat P_\sh \bm h,
	\qquad
	\bm V_{\rm post} = \widehat P_\ob \bm V_{\rm post} \widehat P_\ob + \widehat P_\sh \bm V_{\rm post} \widehat P_\sh,
\eeq
which also apply to the ground truth, $\bar{\bm\epsilon} = \widehat P_\ob \bar{\bm\epsilon} + \widehat P_\sh \bar{\bm\epsilon}$,
and to the prior, $\bm V_0 = \eta_H^2 (\widehat P_\ob + \widehat P_\sh)$.
For each part, we define the one-standard-deviation band of a pixel $\epsilon_m$ by the diagonal element of the projected covariance,
\beq[eq:proj-width]
	(\sigma^{\ob}_m)^2 = \bigpp{ \widehat P_\ob \bm V_{\rm post} \widehat P_\ob }_{mm},
	\qquad
	(\sigma^{\sh}_m)^2 = \bigpp{ \widehat P_\sh \bm V_{\rm post} \widehat P_\sh }_{mm},
\eeq
and similarly the projected prior widths $\eta_H [(\widehat P_\ob)_{mm}]^{1/2}$ and $\eta_H [(\widehat P_\sh)_{mm}]^{1/2}$,
where the repeated $m$ indices are not summed.
By \eq{eq:post-split}, these variances add up to the full variances, 
\beq
	(\sigma^{\rm post}_m)^2 = (\sigma^{\ob}_m)^2 + (\sigma^{\sh}_m)^2,
	\qquad
	\eta_H^2 = \eta_H^2 (\widehat P_\ob)_{mm} + \eta_H^2 (\widehat P_\sh)_{mm}.
\eeq

From \eqs{eq:cov-z-g-svd}{eq:dev-svd}, the two parts behave in opposite ways.
In the shadow GPD space, $\lambda_k^2 \leq 0.01$, so that
\beq
	\widehat P_\sh \bm h \approx \bm 0,
	\qquad
	\widehat P_\sh \bm V_{\rm post} \widehat P_\sh 
		= \eta_H^2 \sum_{k > n_\ob} \frac{ \bm v^{(k)} \bm v^{(k)\T} }{ 1 + \lambda_k^2 } 
		\approx \eta_H^2 \, \widehat P_\sh.
\eeq
which means that the data leave both the mean and the width at the prior,
independently of the data precision.
For the well constrained directions of the observed subspace, $\lambda_k \gg 1$, we have
\beq
	\widehat P_\ob (\bm h - \bar{\bm \epsilon}) 
	\approx 0,
	\qquad
	\widehat P_\ob \bm V_{\rm post} \widehat P_\ob
	\approx \eta_H^2 \sum_{k = 1}^{n_\ob} \frac{\bm v^{(k)} \, \bm v^{(k)\T}}{ \lambda_k^2 },
\eeq
so that the mean approaches the ground truth 
and the uncertainty scales as $r$ independently of $\eta_H$.
Therefore, a well recovered posterior mean requires the truth to lie mostly along the well constrained directions,
while small pointwise uncertainties require a small shadow subspace.

We apply this decomposition to each individual inversion, as shown
in \fig{fig:gpd-level-subspaces} at $\xi = 0.2$, the full posterior of which is shown in \fig{fig:gpd-level-bands-shadow}.
For the ``DVCS'' inversion (left column), 
the observed subspace consists of two directions per slice localized at $x = \xi$.
Its projected prior band is a narrow spike there, the projected truth is a spike of the same shape,
and the projected posterior mean reproduces it with a negligible band.
The shadow part carries the rest of the truth, about the full GPD away from the ridge,
while its posterior mean vanishes and its band coincides with the prior band.
The full posterior is the sum of the two parts:
the mean follows the truth only at the ridge, 
and the band is at the prior width except for a dip at $x = \xi$, where $(\widehat P_\sh)_{mm}$ is small.
The evolution effect (middle column) extends the observed subspace away from the ridge,
but the projected truth in the shadow row remains of the size of the GPD.
With the NSIs (right column), the projected truth in the observed subspace is almost the full truth,
and the projected posterior mean follows it within the projected band,
whereas the shadow row has a vanishing truth, a vanishing mean, and a band at the prior width.
Hence the accurate posterior mean and the wide pointwise bands in \fig{fig:gpd-level-bands-shadow} come from different subspaces:
the mean from the observed one and the bands from the shadow one.

\subsubsection{Correlations and predictions}
\label{sec:results-gpd-corr}

\begin{figure*}[t]
    \centering
    \includegraphics[scale=0.95]{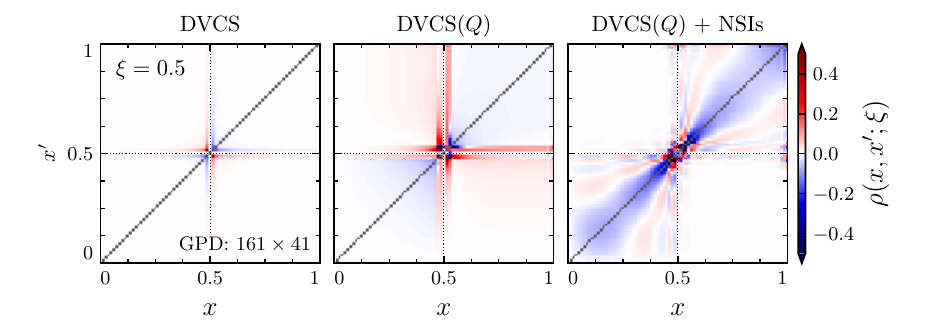}
    \vspace*{-0.5cm}
    \caption{Posterior correlations $\rho(x, x'; \xi)$ in \eq{eq:cor-g} between the GPD pixels at $\xi = 0.5$
    for the inversion at the GPD level with $r = 0.01$ and $\eta_H = 10$,
    for the three combinations of pseudodata (panels).
    The diagonal ($\rho = 1$) is shown in gray, and the color range is limited to $|\rho| \leq 0.5$.
    The dotted lines mark $x = \xi$ and $x' = \xi$.}
\label{fig:gpd-level-corr}
\end{figure*}

\begin{figure*}[h]
    \centering
    \includegraphics[scale=0.95]{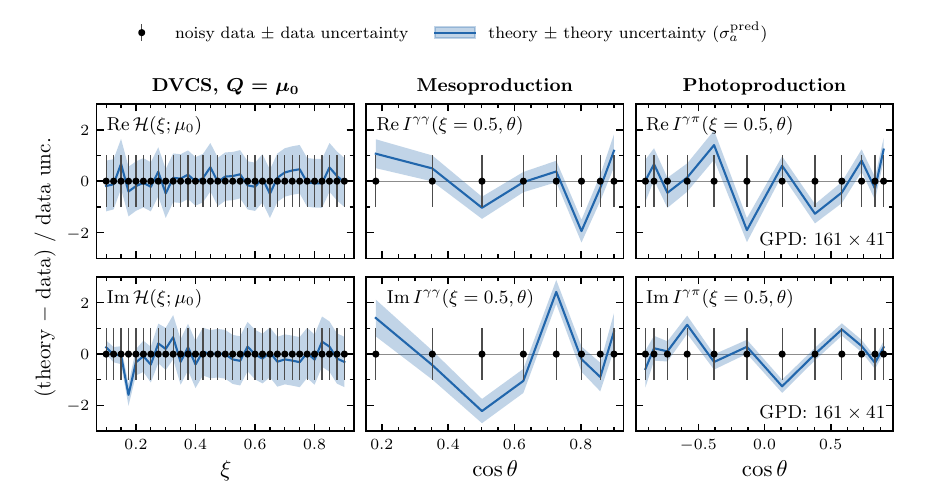}
    \vspace*{-0.5cm}
    \caption{Data-theory comparison for the noisy ``DVCS$(Q)$ + NSIs'' inversion in \fig{fig:gpd-level-maps}(b),
    normalized by the data uncertainty with $r = 0.01$.
    The theory uncertainties $\sigma^{\rm pred}_a$ are computed 
    with the full posterior covariance matrix as in \eq{eq:pred-unc}.}
\label{fig:gpd-level-pred-obs}
\end{figure*}

Although the pointwise uncertainty is dominated by the shadow GPDs, 
they hardly contribute to the off-diagonal elements of the posterior covariance,
which are determined by the constrained directions,
as argued below \eq{eq:cov-z-g-svd-2}.
Since the $\bm J$ here is block diagonal in $\xi$, the pixels at different $\xi$ are uncorrelated,
and we concentrate on the correlations between the pixels at the same $\xi$,
\beq
	\rho(x, x'; \xi)
	= \frac{\Bigvv{ \bigbb{ H^{u+}(x, \xi) - \vv{H^{u+}(x, \xi)} } \bigbb{ H^{u+}(x', \xi) - \vv{H^{u+}(x', \xi)} } }}{
		\sqrt{ \Bigvv{ \bigbb{ H^{u+}(x, \xi) - \vv{H^{u+}(x, \xi)} }^2 } } \, \sqrt{ \Bigvv{ \bigbb{ H^{u+}(x', \xi) - \vv{H^{u+}(x', \xi)} }^2 } }
	}.
\eeq
It is shown in \fig{fig:gpd-level-corr} at $\xi = 0.5$ for each inversion.
Clearly, the correlations mostly concentrate around the diagonal line $x \simeq x'$,
due to the local feature of the well constrained singular directions as illustrated in \fig{fig:gpd-level-directions},
and their oscillation patterns also agree with each other.
The correlations are strongest near the ridge $x, x' \sim \xi$.
With the NSIs, the correlations extend further from the diagonal,
indicating more constraints on the $x$ dependence.
These correlations stay within the DGLAP and the ERBL regions individually, with little correlation between them.

The correlations are essential for the uncertainties of any quantity derived from the GPD.
To illustrate this, we make predictions for the observables from the posterior of the replica of noisy ``DVCS$(Q)$ + NSIs'' pseudodata
in \fig{fig:gpd-level-maps}(b) and compare them with that replica in \fig{fig:gpd-level-pred-obs}.
In units of the data uncertainties, the covariance of the predictions is
\begin{align}
	\vv{ (\bm\Sigma^{-1/2} \delta \bm T) \, (\bm\Sigma^{-1/2} \delta \bm T)^\T }
	&= \bm\Sigma^{-1/2} \bm A \vv{ (\bm \epsilon - \bm h) (\bm \epsilon - \bm h)^\T } \bm A^\T \bm\Sigma^{-1/2} 
	\nn\\
	&= \bm\Sigma^{-1/2} \bm A \bm V_{\rm post} \bm A^\T \bm\Sigma^{-1/2}
	= \bm J \bm V_z \bm J^\T
	= \sum_{k = 1}^{r_A} \frac{ \lambda_k^2 }{ 1 + \lambda_k^2 } \, \bm u^{(k)} \bm u^{(k)\T},
\label{eq:pred-svd}
\end{align}
whose eigenvalues are below $1$.
Hence the predicted uncertainty $\sigma^{\rm pred}_a$ 
never exceeds the data uncertainty $\sqrt{\Sigma_{aa}}$,
\beq[eq:pred-unc]
	\sigma^{\rm pred}_a = \sqrt{ (\bm A \bm V_{\rm post} \bm A^\T)_{aa} }
		\leq \sqrt{\Sigma_{aa}},
\eeq
as seen from the bands in \fig{fig:gpd-level-pred-obs}.
This does not depend on the noise.
Importantly, only the constrained directions enter \eq{eq:pred-svd}; 
since $\bm A \widehat P_\sh \approx \bm 0$,
the shadow part of the posterior, which dominates the pointwise widths, is invisible in the observables.
The correlations induced by the well constrained directions thus 
play a crucial role in making predictions for observables.
Otherwise, if only the diagonal elements of $\bm V_{\rm post}$ are kept,
the variances $\eta_H^2 (V_z)_{mm} \approx \eta_H^2$ enter the predictions 
without the off-diagonal elements that cancel them,
and greatly overestimate the uncertainties,
far beyond the data uncertainties (by up to three orders of magnitude).
Therefore, the shadow GPDs cancel in the observables only through the correlations among the pixels,
and any quantity derived from the GPD pixels must be propagated with the full posterior covariance.
This is a major advantage of the pixelation approach, which gives the full posterior covariance exactly.

\subsubsection{Predictions for other observables}
\label{sssec:pred-moments}

The large pointwise uncertainty of the posterior GPD
severely limits the tomographic imaging of the proton,
which requires precise knowledge of the GPD itself.
But many other quantities of interest are related to the GPD
via convolutions of the form in \eq{eq:gpd-integral},
in which most of the shadow GPDs in the posterior distribution cancel.
Thus, the GPD need not be fully determined to make unambiguous predictions.

As an example, we consider the lowest two nonzero $x$ moments of the GPD,
\begin{align}
     H(n, \xi) = \int_{-1}^{1}\dd{x} x^{n-1} H^{u+}(x, \xi; \mu_0),
     \qquad n=2, \, 4,
\label{eq:moments-n}
\end{align}
where we have fixed the scale to $\mu_0$ and only the even moments are nonzero for the C-even GPD component.
By polynomiality in \eq{eq:gpd-x-moments}, $H(2, \xi)$ is constant and $H(4, \xi)$ is quadratic in $\xi$,
and their coefficients, the generalized form factors, can be computed in lattice QCD~\cite{Gockeler:2003jfa, Hagler:2003jd, LHPC:2007blg, Alexandrou:2011nr, Alexandrou:2013joa, Bali:2018zgl, Alexandrou:2019ali, Alexandrou:2021bbo, Bhattacharya:2024wtg, HadStruc:2024rix, Gao:2025inf}.
With their $t$ dependence, the generalized form factors play an important role in deciphering the mechanical properties of the proton~\cite{Ji:1994av, Ji:1995sv, Ji:1996ek, Polyakov:2002yz, Lorce:2017xzd, Polyakov:2018zvc, Burkert:2018bqq, Kumericki:2019ddg, Shanahan:2018nnv, Lorce:2018egm, Metz:2020vxd, Dutrieux:2021nlz, Freese:2021czn, Burkert:2023wzr, Hackett:2023rif}.
Here, we examine how well the posterior GPDs inverted from the three sets of pseudodata can predict these moments.

Following \eq{eq:obs-mat-prod},
the moments can be represented in the pixelation framework as
\beq[eq:moments-mat-prod]
    H(n, \xi)
    = \sum_{i, j} C^{{\rm M}, (p)}_{ij}(n, \xi) \,
        H^u_{ij}(\mu_0),
\eeq
with the observable matrix,
\begin{align}
    C^{{\rm M}, (p)}_{ij}(n, \xi)
    &= \int_{-1}^1 \dd{x} G^{(p)}_{ij}(x, \xi) \bb{
        x^{n-1}
        - (-x)^{n-1}
    }
    = 2 \, \Xi^{(p)}_j(\xi)
        \int_{-1}^1 \dd{x} X^{(p)}_i(x) \, x^{n-1}.
\label{eq:obs-mat-moments}
\end{align}
The predicted moments are then $\bm C^{\rm M} \bm h$, 
which is a vector in the space of $n$ and $\xi$,
with the covariance matrix $\bm C^{\rm M} \bm V_{\rm post} (\bm C^{{\rm M}})^{\T}$.
For the GPD-level inversion here, the pixels at different $\xi$ are uncorrelated, so are the predicted moments at different $\xi$,
and polynomiality is not guaranteed.

\begin{figure*}[t]
    \centering
    \includegraphics[scale=1.0]{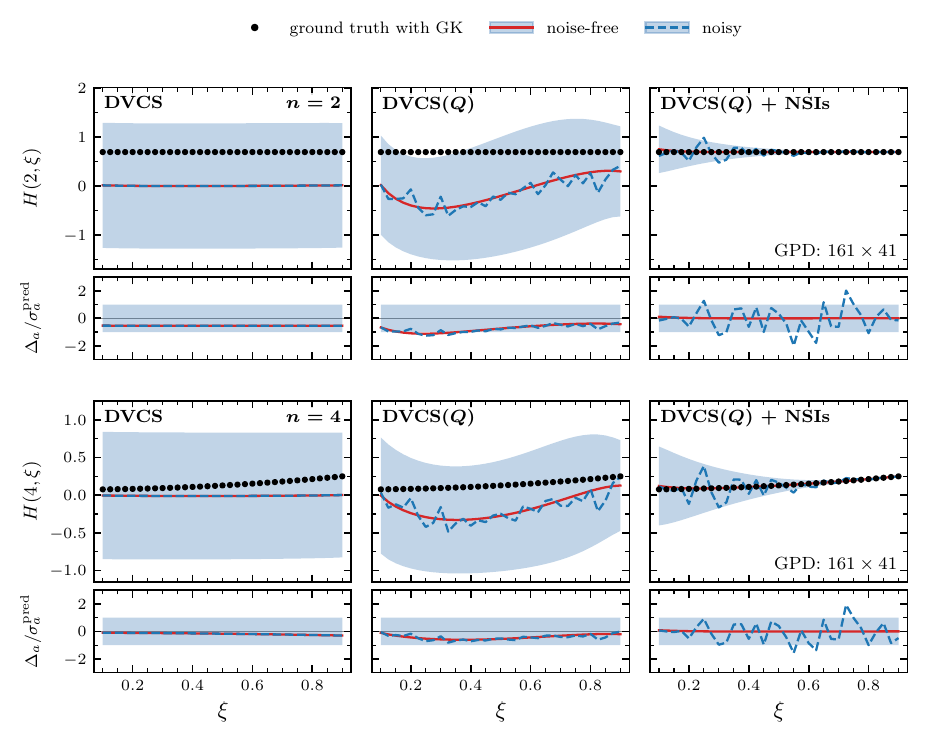}
    \caption{Predictions of the $x$ moments $H(2, \xi)$ (upper row) and $H(4, \xi)$ (lower row) at $\mu_0$ from the posteriors inverted at the GPD level (shown in \fig{fig:gpd-level-maps}), for the three combinations of pseudodata, with (blue dashed curves) and without (red solid curves) noise.
    The theory uncertainty bands (blue shades) are obtained from the full posterior covariance, 
    	$\sigma_a^{\rm pred} = [(\bm C^{\rm M} \bm V_{\rm post} \bm C^{{\rm M}\T})_{aa}]^{1/2}$,
	and hold for both the noise-free and the noisy predictions.
    The lower panel of each row shows the deviations $\Delta_a$ of the predictions from the ground truth, normalized by the theory uncertainty.}
\label{fig:gpd-level-pred-moments}
\end{figure*}

The results are shown in \fig{fig:gpd-level-pred-moments} for the three inversions with noise-free and noisy pseudodata,
in comparison with the moments evaluated from the ground-truth GK model.
As expected, only the ``DVCS$(Q)$ + NSIs'' inversion gives predictions close to the true values.
Even though the pointwise GPD uncertainties are still large,
as shown in \fig{fig:gpd-level-bands-shadow},
they largely cancel in the moments.
For this to happen, it is crucial to use the full correlations among the GPD pixels,
as in \sec{sec:results-gpd-corr}.
Otherwise, the uncertainty is dominated by the shadow GPDs and greatly overestimated.
This feature remains with noisy pseudodata,
for which the predicted moments fluctuate around the truth within the uncertainty bands.

\subsection{Inversion at DD level: impact of polynomiality}
\label{sec:results-dd}

Next, we add the polynomiality constraint by taking the independent DD pixels $\bm\epsilon' \equiv \{ D^{u+}_{H, kl} \}$ as the free parameters,
so that the GPD given by $\bm\epsilon = \bm R \bm\epsilon'$ satisfies polynomiality by construction.
We use the prior $\mathcal{N}(0, \eta_D^2)$ for each DD pixel independently, i.e., $\bm V_0' = \eta_D^2 \mathbf{1}$,
where $\eta_D$ is chosen individually for each given DD grid 
such that the median prior width of the resultant GPD pixels with $x > 0$ equals $\eta_H = 10$, 
which gives $\eta_D = 63.4$, $76.0$, and $90.8$ for $N_u = 81$, $115$, and $163$, respectively,
so that the inversions at the DD and GPD levels start from comparable prior coverage in the GPD space.
As mentioned in \sec{sec:data}, the ground truth of the DD-level inversion is the GK DD on the same DD grid, $\bar{\bm\epsilon}'$,
and the pseudodata are generated from its GPD image $\bm R \bar{\bm\epsilon}'$,
so that each DD inversion is compared with its own ground truth.
We consider the three DD grids in \eq{eq:info-ratio}, $N_u = 81$, $115$, and $163$,
with $n_D = 1640$, $3306$, and $6642$ independent DD pixels, respectively, compared with the $3280$ GPD pixels.
The main problems we address in this subsection 
are what polynomiality adds compared with the GPD level, and how the results depend on the DD grid.

\subsubsection{Singular value decomposition}
\label{sec:results-dd-svd}

As formulated in \sec{ssec:bayes-dd}, the DD-level inversion is the GPD-level one 
with $\sqrt{\bm V_0}$ replaced by $\bm S = \bm R \sqrt{\bm V_0'} = \eta_D \bm R$.
In the GPD space, the posterior mean and covariance are $\bm h = \bm S \bm\mu_{z'}$ and
\beq[eq:post-dd-svd]
	\bm V_{\rm post} = \bm S \bm V_{z'} \bm S^\T
	= \bm S \bm S^\T - \sum_{k = 1}^{r_{AR}} \frac{\lambda_k'^2}{1 + \lambda_k'^2} \, (\bm S \bm v'^{(k)}) (\bm S \bm v'^{(k)})^\T,
\eeq
where $\lambda_k'$ and $\bm v'^{(k)}$ are the singular values and right singular vectors of $\bm J'$ in \eq{eq:J-dd},
whose number is now given by the rank $r_{AR}$ of $\bm A \bm R$.
Compared with \eq{eq:cov-z-g-svd-2}, the uncorrelated prior $\eta_H^2 \mathbf{1}$ is replaced by the prior $\bm S \bm S^\T$ induced by the DD,
which already correlates the GPD pixels, including those at different $\xi$, before any data are included.
Hence, two SVDs are needed to understand the DD-level inversion.
The SVD of $\bm S$ characterizes polynomiality alone,
and the SVD of $\bm J'$ gives the constraints of the data on top of it.
Since the data enter only through $\bm J'$, \eq{eq:chi2-svd} for the $\chi^2$ holds exactly with $\lambda_k$ replaced by $\lambda_k'$ and $r_A$ by $r_{AR}$.
The deviation of the posterior mean from the truth $\bar{\bm\epsilon} = \bm S \bar{\bm z}'$ becomes
\beq[eq:dev-dd]
	\bm h - \bar{\bm\epsilon}
	= - \sum_{k} \frac{ \bar z'_k }{ 1 + \lambda_k'^2 } \, \bm S \bm v'^{(k)}
	+ \sum_{k} \frac{ \lambda_k' \, \delta w_k' }{ 1 + \lambda_k'^2 } \, \bm S \bm v'^{(k)},
\eeq
with $\bar z'_k = \bm v'^{(k)\T} \bar{\bm z}'$ and $\delta w_k' = \bm u'^{(k)\T} \delta\bm w$, where $\bm u'^{(k)}$ are the left singular vectors of $\bm J'$.
The noise term contributes $\sum_k \lambda_k'^2 / (1 + \lambda_k'^2)^2 \, |\bm S \bm v'^{(k)}|^2$ to $\vv{|\bm h - \bar{\bm\epsilon}|^2}$,
as in \eq{eq:deltaH-svd-noise} but weighted by the GPD-space norm of each direction.
The DD directions fall into three classes:
\begin{itemize}
\item those constrained by the data, with $\lambda_k' > 0.1$ and guaranteed to have $\bm S \bm v'^{(k)} \neq \bm 0$;
\item the shadow directions with $\lambda_k' \leq 0.1$ but $\bm S \bm v'^{(k)} \neq \bm 0$, whose images are the shadow GPDs at the DD level;
and
\item those in the null space of $\bm R$, with $\bm S \bm v'^{(k)} = \bm 0$, which change neither the GPD nor the data and are irrelevant.
\end{itemize}

\begin{figure}[t]
    \centering
    \includegraphics[scale=0.95]{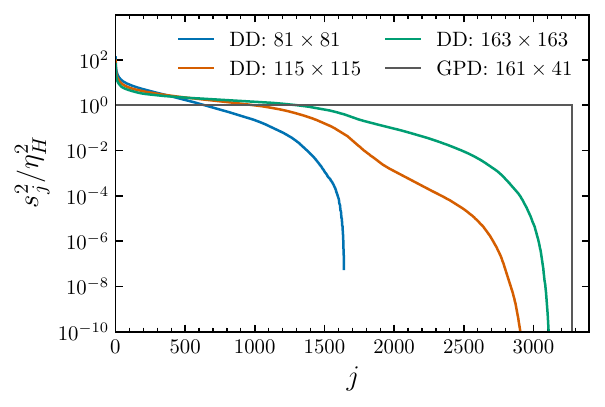}
    \vspace*{-0.3cm}
    \caption{Squared singular values $s_j^2$ of $\bm S = \eta_D \bm R$ (prior variances of the GPD along its singular directions), in units of $\eta_H^2$, for the DD grids with $N_u = 81$, $115$, and $163$.
    The horizontal line marks the GPD-level prior, $\eta_H^2$, for all the $3280$ GPD pixels.}
\label{fig:dd-level-prior-spectrum}
\end{figure}

Figure~\ref{fig:dd-level-prior-spectrum} shows the prior variances along the singular directions of $\bm S$.
Unlike the flat GPD-level prior, they decrease continuously and drop sharply beyond a number of directions set by the DD grid,
e.g., half of the GPD directions have zero prior width for $N_u = 81$.
Polynomiality thus suppresses a large part of the GPD space, more strongly for coarser DD grids,
and puts more prior variance on the remaining directions.
\footnote{In this sense, there is no unambiguous quantitative comparison between the DD-level and GPD-level inversions.}
It does not, however, specifically remove the shadow GPDs of the GPD-level inversion,
which are determined by the observables.

\begin{figure}[t]
    \centering
    \includegraphics[width=0.9\textwidth]{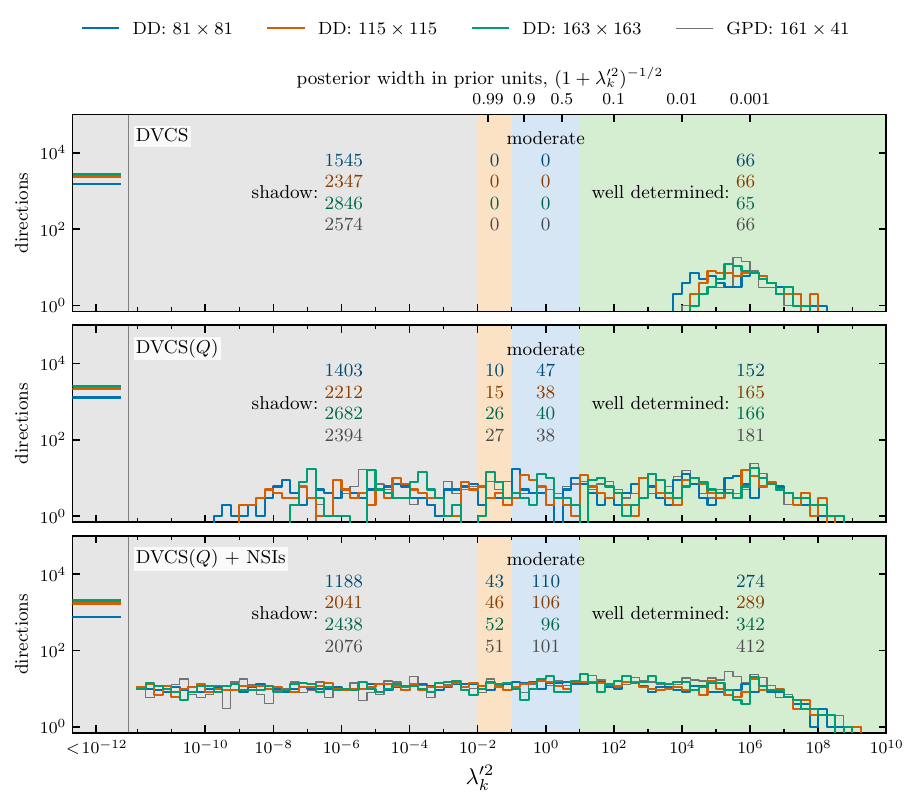}
    \vspace*{-0.3cm}
    \caption{As in \fig{fig:gpd-level-svd-spectrum}, but for the squared singular values $\lambda_k'^2$ of $\bm J'$ in \eq{eq:J-dd} on the DD grids with $N_u = 81$, $115$, and $163$ (colored histograms), compared with the GPD level (gray). The directions in the null space of $\bm R$ are not counted. The numbers of directions in each region are given for $N_u = 81$, $115$, $163$, and the GPD level (top to bottom), in the colors of the corresponding histograms.}
\label{fig:dd-level-svd-spectrum}
\end{figure}

The singular values of $\bm J'$ are shown in \fig{fig:dd-level-svd-spectrum}.
Since part of the GPD space is already suppressed by the prior, the data constrain fewer directions than at the GPD level,
while the moderately constrained ones are about as many.
The spectra hardly depend on $N_u$ in shape;
finer grids have more well constrained directions with the NSIs and more shadow directions.
The main effect of polynomiality is instead that each image $\bm S \bm v'^{(k)}$ spans many $\xi$ slices,
so that a data point at one $\xi$ constrains the GPD at all $\xi$.
For example, the DVCS integrals still constrain about one direction per data point,
but these are no longer localized at single $\xi$ slices.

\subsubsection{Quality of the inversion}
\label{sec:results-dd-mean}

\begin{figure*}[t]
    \centering
    \includegraphics[scale=1.0]{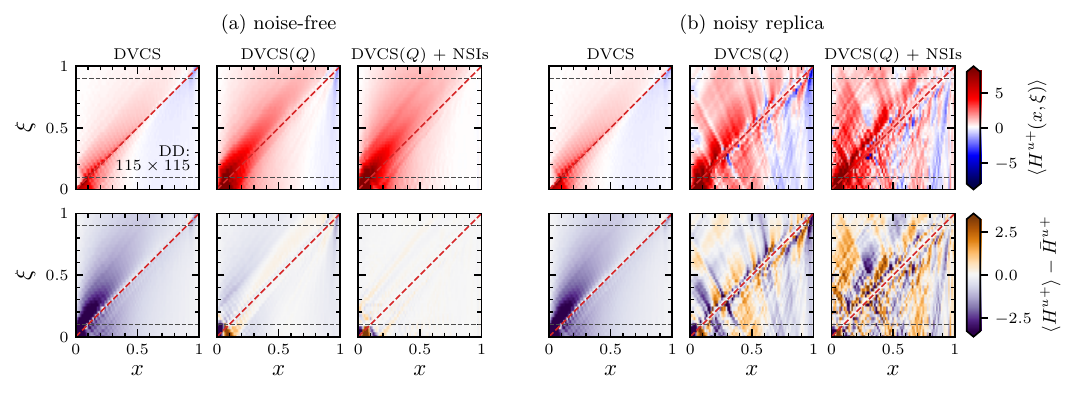}
    \vspace*{-0.7cm}
    \caption{As in \fig{fig:gpd-level-maps}, but for the inversion at the DD level on the equal-information grid $N_u \times N_v = 115 \times 115$, displayed on the GPD grid through $\bm\epsilon = \bm R \bm\epsilon'$. The ground truth is $\bm R \bar{\bm\epsilon}'$ on the same DD grid. Unlike \fig{fig:gpd-level-maps}, the regions $\xi < 0.1$ and $\xi > 0.9$ are not hatched because polynomiality relates all $\xi$ so that the data constrain the GPD there as well.\\ \\}
\label{fig:dd-level-maps}
\end{figure*}

\begin{figure*}[h]
    \centering
    \includegraphics[scale=1.15]{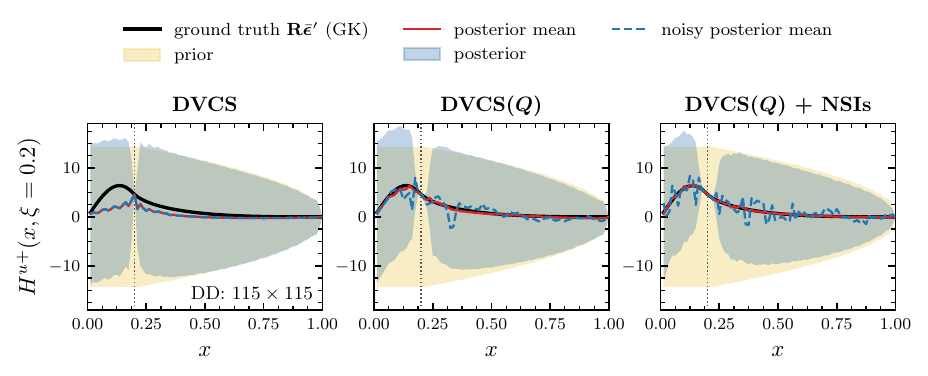}
    \vspace*{-0.7cm}
    \caption{Same as \fig{fig:gpd-level-bands-shadow}, but for the inversion at the DD level on the equal-information grid. The pale-yellow bands show the prior induced on the GPD pixels, $\pm [(\bm S \bm S^\T)_{mm}]^{1/2}$, and the black curves the ground truth $\bm R \bar{\bm\epsilon}'$ from the GK model. \\}
\label{fig:dd-level-bands-shadow}
\end{figure*}

We first discuss the noise-free inversion for the equal-information grid with $N_u = 115$, whose $\delta_H$ and $\chi^2$ are given in the fifth column of Table~\ref{tab:results-chi2}.
The posterior means in the $(x, \xi)$ space are shown in \fig{fig:dd-level-maps}(a) and in the $x$ space with pointwise uncertainty bands in \fig{fig:dd-level-bands-shadow}.
Due to the Radon transform, the prior pointwise GPD uncertainty depends on $\xi$, and is larger at small $x$ than at large $x$.
As at the GPD level, the posterior band stays close to the prior width except near $x = \xi$.
But now the DVCS integrals alone also constrain the GPD away from the ridge due to the polynomiality constraint, reducing $\delta_H$ from $98\%$ to $67\%$.
Many shadow directions of the GPD level have no counterpart at the DD level, being suppressed by polynomiality, and each new shadow GPD spans many $\xi$ slices with an amplitude that depends strongly on $\xi$.
That also makes the evolution much more effective in the ``DVCS$(Q)$'' inversion, which now gives $\delta_H = 19\%$, compared with $88\%$ at the GPD level.
In contrast, for the ``DVCS$(Q)$ + NSIs'' inversion, the $\delta_H$ becomes slightly larger than at the GPD level, as a result of the polynomiality restriction removing some directions that the NSIs could have constrained.
Polynomiality is thus decisive particularly when the data leave most directions unconstrained, but not a substitute for observables sensitive to the $x$ dependence.

This also reflects in the DD grid dependence given in the last three columns of Table~\ref{tab:results-chi2},
where two clearly opposite trends appear.
For the DVCS integrals with or without the evolution, a coarser DD grid gives a smaller $\delta_H$,
since a finer DD grid suppresses fewer GPD directions and leaves more shadow directions,
as shown in \fig{fig:dd-level-prior-spectrum},
which increases the shadow part of the truth GPD.
For the ``DVCS$(Q)$ + NSIs'' data, which constrain many more directions,
the finest grid gives the smallest $\delta_H$, as it restricts the GPD less
and the data constrain more directions.

\begin{figure*}[t]
    \centering
    \includegraphics[scale=1.15]{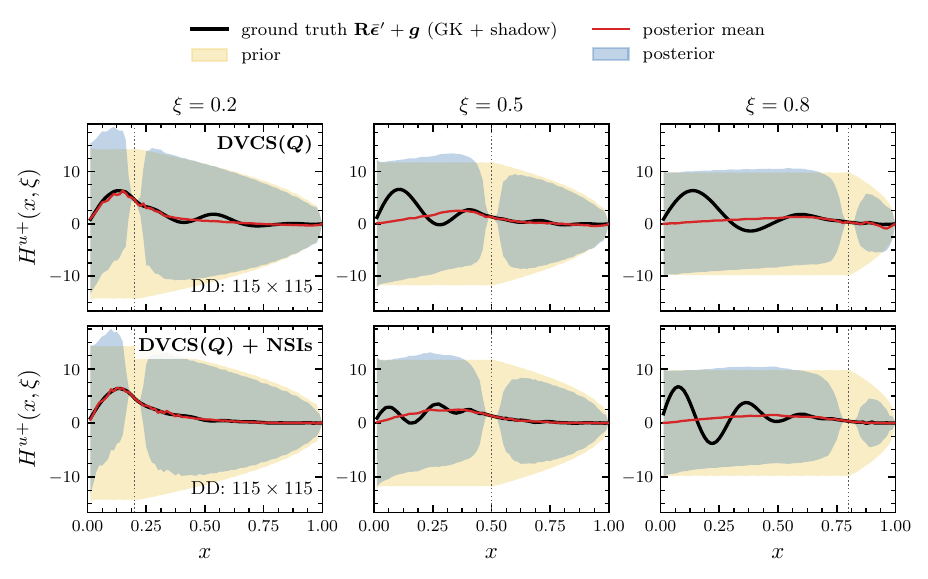}
    \vspace*{-0.5cm}
    \caption{As in \fig{fig:shadow-truth}, but for the inversion at the DD level on the equal-information grid with the ``DVCS$(Q)$'' (upper row) and ``DVCS$(Q)$ + NSIs'' (lower row) pseudodata, with the truth $\bm R \bar{\bm\epsilon}' + \bm g$, where $\bm g$ is a smooth shadow GPD at the DD level for each data set. \\ \\ }
\label{fig:shadow-truth-dd}
\end{figure*}

As the inversion at the GPD level,
the fact that here both the ``DVCS$(Q)$'' and ``DVCS$(Q)$ + NSIs'' inversions 
recover the ground-truth in their posterior means to a good extent
depends on the GK model we start with.
The shadow GPDs remain as the directions invisible to the data.
Similarly to the GPD-level study in \fig{fig:shadow-truth}, 
we test this by adding a smooth shadow GPD $\bm g$ at the DD level to the ground truth, $\bm R \bar{\bm\epsilon}' + \bm g$,
and repeat the two inversions, with the result shown in \fig{fig:shadow-truth-dd}.
In both cases, the posterior misses the shadow completely 
and still gives the same result as for the unmodified ground truth.

\begin{figure*}[h]
    \centering
    \includegraphics[scale=1.15]{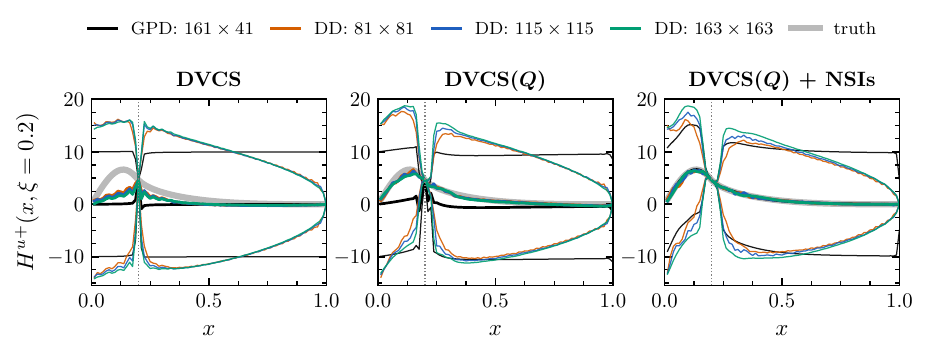}
    \vspace*{-0.5cm}
    \caption{Comparison of the inversions at the DD level on the half- ($N_u = 81$), equal- ($115$), and double-information ($163$) grids, and at the GPD level, at $\xi = 0.2$ for the three combinations of pseudodata, with $r = 0.01$. The thick curves show the posterior means, and the thin curves of the same colors mark the one-standard-deviation bands. The gray curves show the ground truth, whose differences between the grids are invisible on this scale.}
\label{fig:dd-grid-comparison}
\end{figure*}

When noise on the data is included, 
the posterior mean develops fluctuations,
as shown in \fig{fig:dd-level-maps}(b) in the $(x, \xi)$ space
and in \fig{fig:dd-level-bands-shadow} for the slice $\xi = 0.2$.
This is governed by the second term of \eq{eq:dev-dd},
which is proportional to $\bm S \bm v'^{(k)}$ and spreads over many $\xi$ slices,
in contrast to the noise within single $\xi$ slices at the GPD level in \fig{fig:gpd-level-maps}(b).
The inversion quality is given in Table~\ref{tab:results-chi2-noisy}.

Comparing this to Table~\ref{tab:results-chi2}, we find that most of the polynomiality improvement is lost especially for ``DVCS$(Q)$'', while the ``DVCS'' inversion results are hardly affected. 
Also, compared with the GPD-level inversion, the ``DVCS$(Q)$ + NSIs'' data now give a larger $\delta_H$, while the data without the NSIs remain better. 
This is because, by \eq{eq:dev-dd}, the noise is again carried by the moderately constrained directions, with about the same total noise factor $\sum_k \lambda_k'^2 / (1 + \lambda_k'^2)^2$ as at the GPD level, as indicated by \fig{fig:dd-level-svd-spectrum}, but each direction now enters with the extra weight $|\bm S \bm v'^{(k)}|^2$, 
which on average exceeds $\eta_H^2$ with the NSIs.
The Radon transform thus enhances the moderately constrained directions of the NSIs.
As seen from the last three columns of Table~\ref{tab:results-chi2-noisy},
this effect weakens on finer DD grids,
which spread the prior variance over more directions (cf.~\fig{fig:dd-level-prior-spectrum}).

Overall, however, the grid dependence is very moderate.
Among the three grids, $\delta_H$ of each data set varies by at most a factor of about $1.5$, with or without noise,
much less than between the data sets or between the GPD and DD levels.
This is more clearly shown in \fig{fig:dd-grid-comparison}
for the posterior means and bands of the three grids,
which agree with each other to a good extent.
The bands only slowly grow wider as $N_u$ increases, approaching those of the GPD level, as polynomiality restricts the GPD less.

\clearpage
\subsubsection{Uncertainties and correlations}
\label{sec:results-dd-unc}

\begin{figure*}[t]
    \centering
    \includegraphics[scale=0.9]{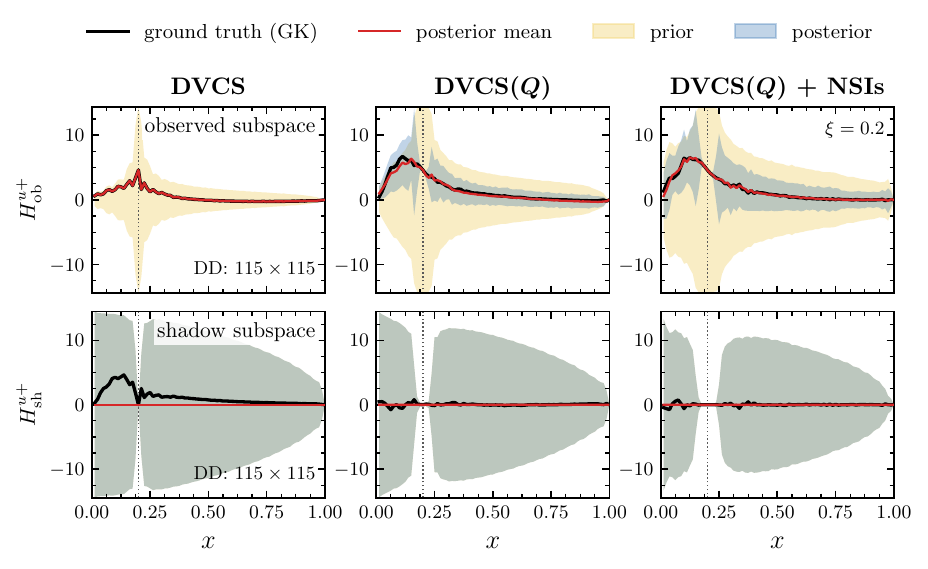}
    \vspace*{-0.3cm}
    \caption{As in \fig{fig:gpd-level-subspaces}, but for the inversion at the DD level on the equal-information grid. \\}
\label{fig:dd-level-subspaces}
\end{figure*}

As at the GPD level in \sec{sec:results-gpd-proj}, we split the posterior into its observed and shadow parts, now defined in the DD space by the SVD of $\bm J'$: $\widehat P'_\ob = \sum_{\lambda_k'^2 > 0.01} \bm v'^{(k)} \bm v'^{(k)\T}$ and $\widehat P'_\sh = \mathbf{1} - \widehat P'_\ob$, the latter including the null space of $\bm R$. Since they commute with $\bm V_{z'}$, mapping to the GPD pixels by $\bm S$ gives
\beq[eq:post-split-dd]
	\bm h = \bm S \widehat P'_\ob \bm\mu_{z'} + \bm S \widehat P'_\sh \bm\mu_{z'},
	\qquad
	\bm V_{\rm post} = \bm S \widehat P'_\ob \bm V_{z'} \widehat P'_\ob \bm S^\T + \bm S \widehat P'_\sh \bm V_{z'} \widehat P'_\sh \bm S^\T,
\eeq
and similarly $\bm R \bar{\bm\epsilon}' = \bm S \widehat P'_\ob \bar{\bm z}' + \bm S \widehat P'_\sh \bar{\bm z}'$ for the ground truth and $\bm S \bm S^\T = \bm S \widehat P'_\ob \bm S^\T + \bm S \widehat P'_\sh \bm S^\T$ for the prior GPD. 
The pixel variances again add without cross terms, although $\bm S$ does not keep the two parts orthogonal on the GPD pixels.
The results are shown in \fig{fig:dd-level-subspaces} at $\xi = 0.2$. 
The main difference from the GPD level is that each observed direction spans all $\xi$, so the observed part at one $\xi$ also carries the constraints from the data at the other $\xi$. For the DVCS integrals, the observed part is no longer a spike at $x = \xi$. 
Polynomiality greatly improves the inversion with the evolution, which now places most of the GK model in the observed subspace.
With the NSIs, the performance is similar to that at the GPD level.
As at the GPD level, the posterior mean and the pointwise bands come from different subspaces.
In all cases, the uncertainty band of the shadow part stays at the induced prior width, and is responsible for the large pointwise bands observed in \fig{fig:dd-level-bands-shadow}.

\begin{figure*}[h]
    \centering
    \includegraphics[scale=0.9]{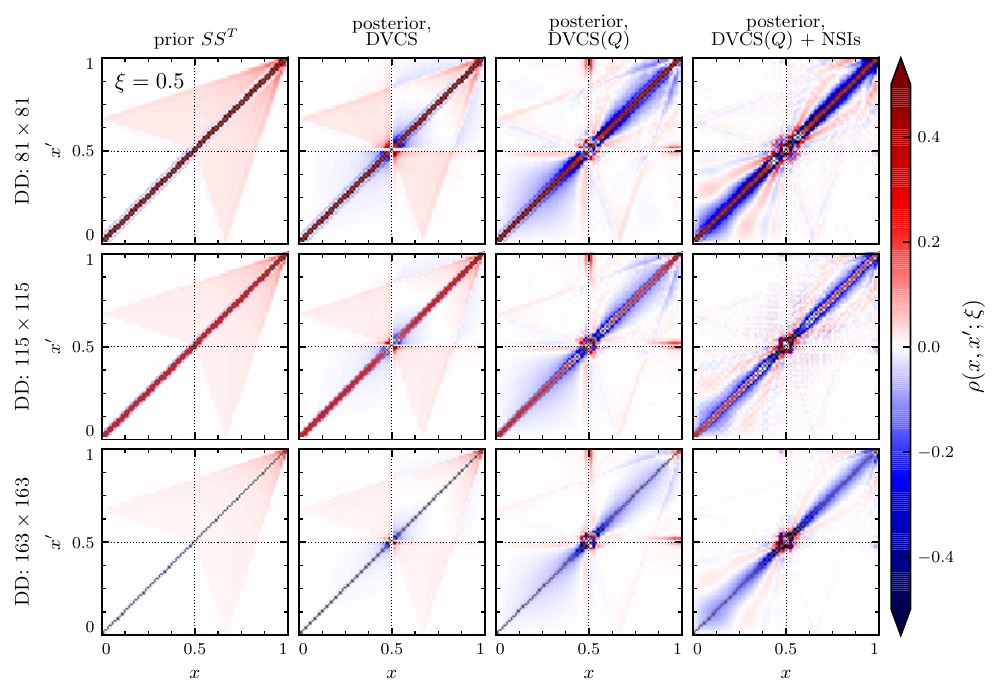}
    \vspace*{-0.3cm}
    \caption{Correlations $\rho(x, x'; \xi)$ between the GPD pixels at $\xi = 0.5$ for the three DD grids (rows). From left to right are correlations in the prior, $\bm S \bm S^\T$, induced by the DD prior and Radon transformation, and in the posteriors for the three combinations of pseudodata. The color range is the same as in \fig{fig:gpd-level-corr}, and the dotted lines mark $x = \xi$ and $x' = \xi$.}
\label{fig:dd-level-corr-grid}
\end{figure*}

The correlations are shown in \fig{fig:dd-level-corr-grid}.
Unlike at the GPD level, the prior $\bm S \bm S^\T$ already correlates the GPD pixels.
Within a $\xi$ slice, these prior correlations are short-ranged, and the data add correlations further away, more so with evolution and further with the NSIs.
All these correlations become weaker on finer grids.
Compared with the GPD level, the correlations for the DVCS integrals with and without evolution are much stronger, whereas those for the ``DVCS$(Q)$ + NSIs'' data have similar sizes but extend further from $x = \xi$.
Polynomiality also induces correlations across $\xi$ (not shown), which come from the prior and are hardly changed by the data; they are strong between neighboring $\xi$ slices and become weaker on finer grids.
These correlations propagate the constraints of each observable across $\xi$, which underlies the improvements of the posterior mean.

\begin{figure*}[htbp]
    \centering
    \includegraphics[scale=1.02]{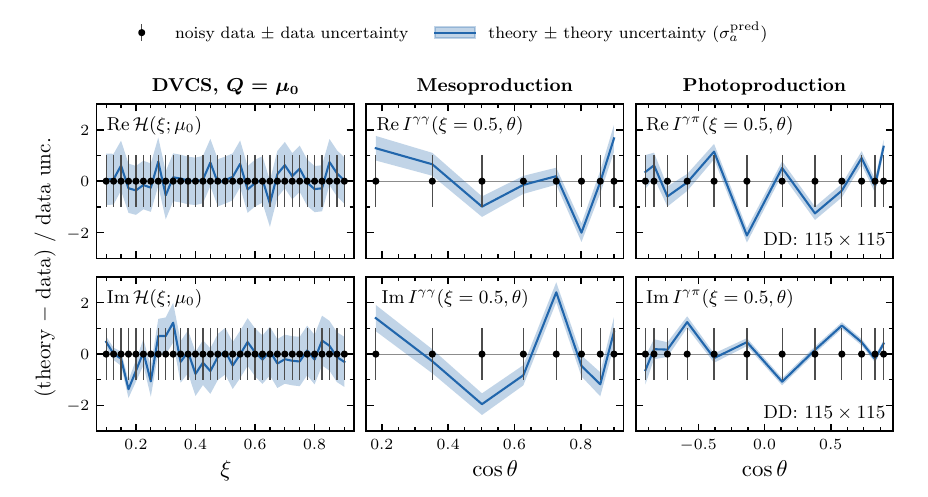}
    \vspace*{-0.3cm}
    \caption{As in \fig{fig:gpd-level-pred-obs}, but for the inversion at the DD level on the equal-information grid. \\}
\label{fig:dd-level-pred-obs}
\end{figure*}

\begin{figure*}[htbp]
    \centering
    \includegraphics[scale=1.02]{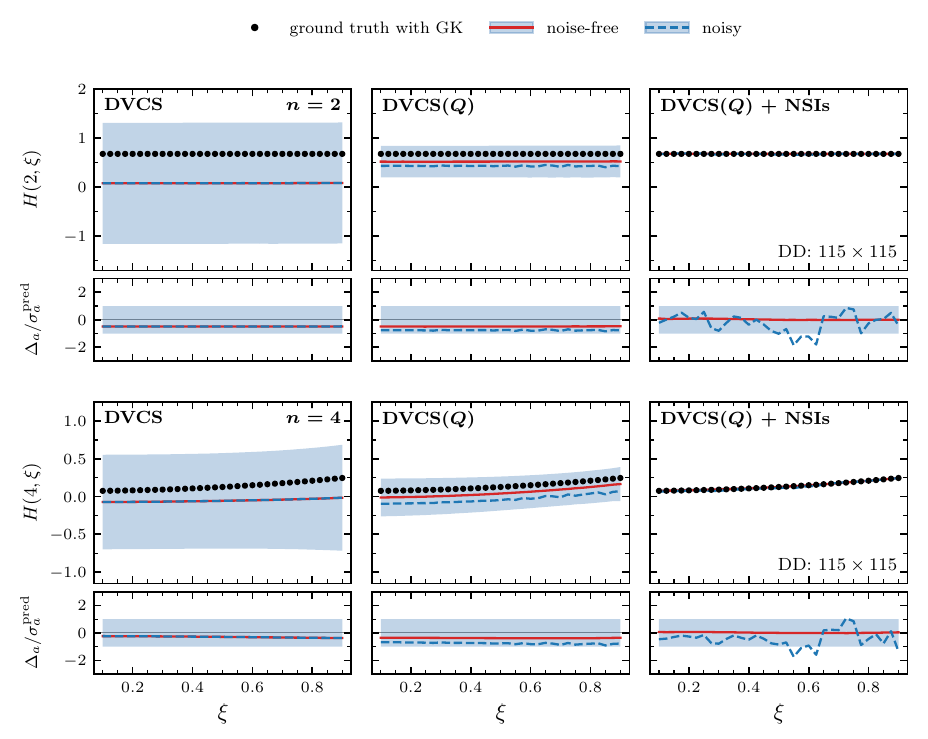}
    \vspace*{-0.3cm}
    \caption{As in \fig{fig:gpd-level-pred-moments}, but for the inversion at the DD level on the equal-information grid.}
\label{fig:dd-level-pred-moments}
\end{figure*}

These correlations are essential in the predictions made of the posterior GPD,
and play an important role of cancelling the shadow GPD contributions.
As an example similar to the GPD-level study in \fig{fig:gpd-level-pred-obs}, 
we show the observables evaluated with the posterior of the DD-level inversions of the noisy ``DVCS$(Q)$ + NSIs'' pseudodata
in \fig{fig:dd-level-pred-obs}, 
which exhibit well defined means and uncertainties.
Had we only included the diagonal elements of the covariance matrix,
the observable band due to GPD uncertainty again exceeds the data uncertainties by up to three orders of magnitude.
Similarly, the $x$ moments computed with the posterior GPD in each DD-level inversion are shown in \fig{fig:dd-level-pred-moments}.
Compared with the GPD-level result in \fig{fig:gpd-level-pred-moments},
the built-in polynomiality now ties the moments at different $\xi$ together,
giving smoother dependence on $\xi$ even for the noisy data. 
In all cases, the deviation from the truth moments lies within the uncertainty band predicted with the full covariance matrix.

\subsection{Dependence on data uncertainty}
\label{sec:results-robust}

Lastly, we repeat the above analyses of the noise-free pseudodata with $r = 0.1$ and $0.2$, keeping the same data points in Table~\ref{tab:pseudodata-summary}, to see how the impact of each set of pseudodata changes with the data precision.
The resulting $\delta_H$ and $\kappa$ are summarized in Table~\ref{tab:results-r}.
The different data sets respond to $r$ in qualitatively different ways.

For the ``DVCS'' inversion, $\delta_H$ is essentially independent of $r$ at both the GPD and DD levels.
Since $\lambda_k \propto \eta_H / r$, the singular values decrease with $r$ but remain in the well constrained range of \fig{fig:gpd-level-svd-spectrum}.
Increasing the uncertainties thus only broadens the posterior along the constrained directions, while the posterior mean, which is determined by the noise-free pseudodata, is barely changed.
The large $\delta_H$ is set by the unconstrained directions, which do not depend on $r$.

In contrast, the constraints from the evolution quickly degrade with $r$,
with both $\kappa$ decreasing and $\delta_H$ increasing.
This happens as the spectrum in \fig{fig:gpd-level-svd-spectrum} shifts to smaller $\lambda_k$, moving well constrained directions into the moderately constrained and shadow ranges, and is more dramatic at the DD level, where the evolution is most effective.
This is expected since the evolution changes the DVCS integrals by no more than $10\%$ in many data points over $Q \in [2, 10]~\GeV$ (cf.~\fig{fig:dvcs-integral}), which becomes comparable to the data uncertainty for $r \gtrsim 0.1$,
so the information beyond the DVCS integrals at a single scale is largely lost.

With the NSIs, the data remain the most constraining at all $r$.
While $\kappa$ decreases with $r$, $\delta_H$ increases only to a few percent at both level.
This holds at both the GPD level and the DD level, and is independent of the DD grid.

\begin{table}[t]
\centering
\caption{Relative deviation $\delta_H$ (in \%) of the posterior mean from the ground truth and the effective number of constraints $\kappa$ for each combination of noise-free pseudodata, for relative uncertainties $r = 0.01$, $0.1$ and $0.2$, and and $p = 3$. For the DD level, $\kappa$ is evaluated in the DD pixel space with the prior width $\eta_D$.}
\label{tab:results-r}
\setlength{\tabcolsep}{8pt}
\begin{tabular}{l|ccc|ccc}
\hline\hline
 & \multicolumn{3}{c}{$\delta_H$ (\%)} & \multicolumn{3}{c}{$\kappa$} \\
Pseudodata & $r = 0.01$ & $0.1$ & $0.2$ & $r = 0.01$ & $0.1$ & $0.2$ \\
\hline
\multicolumn{7}{l}{GPD} \\
\quad DVCS & 97.9 & 97.9 & 97.9 & \phantom{0}66 & \phantom{0}66 & \phantom{0}66 \\
\quad DVCS$(Q)$ & 88.1 & 93.8 & 95.1 & 202 & 157 & 144 \\
\quad DVCS$(Q)$ + NSIs & \phantom{0}1.9 & \phantom{0}4.7 & \phantom{0}5.7 & 465 & 350 & 314 \\
\hline
\multicolumn{7}{l}{DD ($N_u = 81$)} \\
\quad DVCS & 61.6 & 61.6 & 61.7 & \phantom{0}66 & \phantom{0}66 & \phantom{0}66 \\
\quad DVCS$(Q)$ & 15.7 & 37.4 & 42.8 & 171 & 135 & 123 \\
\quad DVCS$(Q)$ + NSIs & \phantom{0}5.3 & \phantom{0}5.7 & \phantom{0}6.4 & 328 & 222 & 193 \\
\hline
\multicolumn{7}{l}{DD ($N_u = 115$)} \\
\quad DVCS & 67.1 & 67.2 & 67.2 & \phantom{0}66 & \phantom{0}66 & \phantom{0}66 \\
\quad DVCS$(Q)$ & 19.5 & 45.5 & 51.8 & 178 & 141 & 129 \\
\quad DVCS$(Q)$ + NSIs & \phantom{0}5.9 & \phantom{0}6.8 & \phantom{0}7.6 & 341 & 233 & 202 \\
\hline
\multicolumn{7}{l}{DD ($N_u = 163$)} \\
\quad DVCS & 72.3 & 72.3 & 72.4 & \phantom{0}65 & \phantom{0}65 & \phantom{0}65 \\
\quad DVCS$(Q)$ & 22.5 & 51.3 & 57.6 & 185 & 145 & 133 \\
\quad DVCS$(Q)$ + NSIs & \phantom{0}3.9 & \phantom{0}5.8 & \phantom{0}6.4 & 392 & 278 & 241 \\
\hline\hline
\end{tabular}
\end{table}

\newpage
\section{Conclusion}
\label{sec:con}

Extracting the full $x$ dependence of GPDs from hard exclusive observables is a nontrivial task and remains a bottleneck of the QCD tomography program.
In this paper, we examine this problem in an ideal scenario by assuming that the basic GPD convolutions have already been extracted with full flavor separation and high precision. 
By constructing the GPD on a discretized $x$-$\xi$ grid, we turn its inversion into a linear Gaussian Bayesian problem, which is exactly solvable and gives the posterior mean and full covariance.
The whole analysis can be formulated with the singular value decomposition, which provides a transparent understanding of the results.

For the various types of observables considered, we find that while the data can always be fitted very well, the underlying GPD cannot be recovered in most cases.
The most important constraints come from the NSIs in the diphoton mesoproduction and photon-pion photoproduction processes.
These bring the posterior mean close to the true GPD in our test case with the sufficiently smooth GK model, but still leave substantial room for shadow GPDs.
With polynomiality imposed, the DVCS data alone with evolution also significantly improve the recovery of the GPD.
However, this relies on the very high precision of the data and rapidly degrades as the data uncertainties increase, whereas the constraints from the NSIs remain stable.

In all cases, the inversion is highly susceptible to the noise in the data.
Once it is included, the posterior mean develops non-negligible fluctuations. 
We find that the impact of the noise depends strongly on the width of the prior, rather than on the precision or the quantity of data.
Its large size here is mostly due to the very loose prior in our study, but without further theoretical input this choice remains largely arbitrary.
Not included in this work are stricter smoothness conditions on the GPD grid beyond the finite element method and polynomiality and, through this, the constraint from the forward limit.
That said, it is clear that more constraints from new physical processes and/or lattice QCD computations are needed to obtain a more complete picture of GPDs and thus the tomographic imaging of the proton.

Leaving aside possible future improvements, we can reflect on how our results should be interpreted at this point.
An advantage of the pixelation method is that the inversion is well defined even when the data do not fix the central GPD, and its result is the full posterior distribution.
Point by point in the $x$-$\xi$ space, the uncertainty of the GPD is dominated by the ``shadow'' degrees of freedom that are not restricted by the data.
However, the posterior also fully captures the correlations among the points, which cancel these degrees of freedom in any observable.
It is these correlations that enable us to compute physically interesting quantities dependent on GPDs such as their $x$ moments, give predictive power to the factorization formulas involving GPDs, as we have verified with data included in the inversion, and allow tests of their universality across different experimental processes, even if the GPDs themselves are not uniquely determined.
This distinctive uncertainty quantification originates from the exclusive nature of GPDs.

Therefore, the fact that the GPD is not uniquely restricted by the currently known observables does not mean that the study of GPDs has no way forward, but rather that we should quantify the uncertainty in a different way.
Instead of the conventional way of assigning a pointwise uncertainty band to a central distribution curve, as we have in \figs{fig:gpd-level-bands-shadow}{fig:dd-level-bands-shadow}, we should supplement it with the full correlations among the $(x, \xi)$ points.
This is sufficient to interpret all current data through the collinear GPD factorization and to make predictions for future data.
At the very least, it allows us to test the collinear factorization to verify the partonic picture underlying exclusive processes, which is nontrivial, still rarely done, and offers more insight into QCD.

\section*{Acknowledgements}
We thank Daniel~Adamiak and Adam~Freese for their collaboration in the early stage of this work, and Ian~Clo\"{e}t, Yuxun Guo, Brandon~Kriesten, Yaohang~Li, Eric~Moffat, and Feng Yuan for helpful discussions and communications. 
This work is supported in part by the U.S. Department of Energy, Office of Science, Office of Nuclear Physics under Contract No. 89243126CSC000213, and the Jefferson Lab LDRD program Projects No.~LD2312 and No.~LD2406. 
The work of N.~S., Z.~Y., and M.~Z.\ at Jefferson Lab was also supported by the DOE, Office of Science, Office of Nuclear Physics in the Early Career Program. 
The work of Z.~Y.\ at Brookhaven National Laboratory is supported by the U.S. Department of Energy under Contract No.~DE-SC0012704.
This work has benefited from interactions within the Quark-Gluon Tomography (QGT) Topical Collaboration funded by the DOE, Office of Science, Office of Nuclear Physics with Award No.~DE-SC0023646.

\bibliography{reference}

\end{document}